\documentclass[12pt,preprint]{aastex}
\usepackage{rotating}
\shorttitle{Multiwavelength Study of the Star-Forming Region S255-S257}
\shortauthors{Ojha et al.}
\newcommand{\msun}{\mbox{\rm $M_{\odot}$}}

\newcommand{\lsun}{\mbox{\rm $L_{\odot}$}~}

\newcommand{\arcs}{\hbox{$^{\prime\prime}$}}

\newcommand{\arcm}{\mbox{$^{\prime}$}}

\newcommand{\into}{\mbox{$\times$~}}

\newcommand{\jhk}{\mbox{$JHK_{\rm s}$}~}
\newcommand{\ks}{\mbox{$K_{\rm s}$~}}
\newcommand{\ksb}{\mbox{$K_{\rm s}$-band~}}

\newcommand{\av}{\mbox{$A_{\rm V}$}}

\newcommand{\hii}{\mbox{H~{\sc ii}~}}

\newcommand{\uchii}{\mbox{UC~H{\sc ii}}}

\newcommand{\hei}{\mbox{He~{\sc i}~}}
\newcommand{\heii}{\mbox{He~{\sc ii}~}}
\newcommand{\mgii}{\mbox{Mg~{\sc ii}~}}
\newcommand{\siii}{\mbox{Si~{\sc ii}~}}
\newcommand{\Siii}{\mbox{Si~{\sc iii}~}}
\newcommand{\siiv}{\mbox{Si~{\sc iv}~}}
\newcommand{\kms} {\hbox{km~s$^{-1}$~}}

\newcommand{\tco}{\hbox{$^{13}$CO}~}

\newcommand{\app}{\mbox{$\approx$}~}
\usepackage{txfonts}
\begin{document}
\title{Star Formation Activity in the Galactic \hii Complex S255-S257}
\author{D. K. Ojha\altaffilmark{1},
M. R. Samal\altaffilmark{2},
A. K. Pandey\altaffilmark{2},
B. C. Bhatt\altaffilmark{3},
S. K. Ghosh\altaffilmark{1},
Saurabh Sharma\altaffilmark{2},
M. Tamura\altaffilmark{4},
V. Mohan\altaffilmark{5} and
I. Zinchenko\altaffilmark{6}
}
\email{ojha@tifr.res.in}
\altaffiltext{1}{Tata Institute of Fundamental Research, Mumbai (Bombay) 400 005, India; ojha@tifr.res.in}
\altaffiltext{2}{Aryabhatta Research Institute of Observational Sciences, Nainital 263 129, India}
\altaffiltext{3}{Indian Institute of Astrophysics, Koramangala, Bangalore 560 034, India}
\altaffiltext{4}{National Astronomical Observatory of Japan, Mitaka, Tokyo 181-8588, Japan}
\altaffiltext{5}{Inter-University Centre for Astronomy and Astrophysics, Pune 411 007, India}
\altaffiltext{6}{Institute of Applied Physics of the Russian Academy of Sciences, Ulyanova 46, 603950 Nizhny Novgorod, Russia}

\begin{abstract}
We present results on the star-formation activity of an optically 
obscured region containing 
an embedded cluster (S255-IR) and molecular gas between two 
evolved \hii regions S255 and S257. We have studied the complex 
using optical, near-infrared (NIR) imaging, optical 
spectroscopy and radio continnum mapping at 15 GHz, along with
{\it Spitzer}-IRAC results. It is found that the main exciting sources of 
the evolved \hii regions S255 and S257 and the compact \hii regions 
associated with S255-IR are of O9.5 - B3 V nature,
consistent with previous observations. Our NIR observations reveal 109 
likely young stellar object (YSO) candidates in an area 
of $\sim$ 4\arcm.9 $\times$ 4\arcm.9 centered on S255-IR, which
include 69 new YSO candidates. To see the global star 
formation, we constructed the $V-I/V$ diagram for 51 optically identified IRAC YSOs  
in an area of $\sim$ 13$\arcm$ $\times$ 13$\arcm$ centered on S255-IR. 
We suggest that these YSOs have an approximate age between 0.1 - 4 Myr, 
indicating a non-coeval star formation. 
Using spectral energy distribution (SED) models, we 
constrained physical properties and evolutionary status of 31 and 16 YSO candidates 
outside and inside the gas ridge, respectively. The models suggest that the 
sources associated within the gas ridge are of younger 
population (mean age $\sim$ 1.2 Myr) than the sources outside the gas ridge 
(mean age $\sim$ 2.5 Myr). The positions of the young sources inside 
the gas ridge at the interface of the \hii regions S255 and S257, 
favor a site of induced star formation. 
\end{abstract}


\section{Introduction}

OB stars have the most dramatic influence on the evolution of
surrounding medium, thereby regulating the star-formation activity
of a complex. However, their evolution in a clustered environment 
and how they interact with the cold interstellar medium and process 
local gas to form  new generation young
stars is still not clear. Ultracompact (UC) \hii regions are 
manifestations of newly
formed massive (O or early B) stars deeply embedded in 
the parental cloud.
The UC \hii regions further evolve into the less obscure stage of compact  
and classical \hii regions.
During the expansion they may sweep low density
material forming dense layers of gas or interact with pre-existing
dense clumps leading to density enhancement. In both the cases, matter at a later stage  
becomes unstable against self gravity and  may lead to new star formation 
(Elmegreen \& Lada 1977). However, the expansion usually takes place 
in non-homogeneous medium and 
the surroundings of the \hii regions are generally rich in molecular chemistry, 
which result in a complicated morphology. Therefore, a detailed study of the star-forming region (SFR) 
at various wavebands is necessary to trace the evolutionary
status of young sources and star-formation scenario of an \hii 
environment.
In this paper, we present results of an optically 
obscured young massive SFR, located between two evolved \hii regions, namely 
S255 and S257 (Sharpless 1959).

The photometric spectral type suggests 
the exciting sources of S255 and S257 are B0V (Moffat et al. 1979) and 
B0.5V (Reid 2008) type of stars. These two \hii regions are itself a part of 
a larger complex consisting of five \hii regions namely S254, S255, S256, 
S257 and S258. The complex is situated at a distance of $\sim$ 2.5 kpc  
(Russeil, Adami, \& Georgelin 2007) and is a part
of Gemini OB association (Huang \& Thaddeus 1986). 
In the present study, we have adopted the distance of  2.5 kpc
for our analysis. However, we also derived distance to the region using spectroscopic
observations that come out to be close to the value of 2.5 kpc (see Section 3).
A cluster S255-IR (also called as S255-2) 
containing infrared (IR)-excess sources (Howard,  Pipher, \& Forrest 1997)
is embedded in the optically obscured region and is also associated with
compact and \uchii~ regions, 
namely  S255-2a, S255-2b,
and S255-2c, whose exciting sources are B1 zero-age-main-sequence (ZAMS) stars
(Snell \& Bally 1986), of which two are optically visible. 
Radio observations also reveal an \uchii~ region $\sim$
1$^{\prime}$ north of S255-2, which is a far-IR source detected by
Jaffe et al. (1984) and is  called S255N. Its total luminosity is 
approximately twice
that of S255-IR (5 $\times$ 10$^4$\lsun; Minier et al. 2005). Sub-millimeter and  millimeter emissions show
large column of gas and dust with sub-structures (Zinchenko, Henning, \& Schreyer 1997;
Klein et al. 2005; Minier et al. 2005; Zinchenko, Caselli, \& Pirogov 2009), elongated 
along the north-south directions
and sandwiched between the two \hii regions  S255 and S257.
Figure 1 shows the H$\alpha$ continuum image,  where individual \hii regions
are marked.  The dark lane between S255 and S257 is the subject of the present study
and is devoid of optical stellar sources. However, this region shows several
signatures of recent star formation such as: IR-excess sources (Howard,  Pipher, \& Forrest 1997;
Miralles et al. 1997), Herbig-Haro objects (Miralles et al. 1997), compact and \uchii~
regions (Snell \& Bally 1986; Kurtz, Churchwell, \& Wood 1994), dense millimeter cores
(Cyganowski, Brogan, \& Hunter 2007) and  masers (Goddi et al. 2007), therefore indicating that
star formation is currently active in the dark lane. Since the dark lane
contains a lot of gas and dust, as found by previous authors,  hereafter we 
refer to this as ``gas ridge'' for further discussion in the paper.

A number of efforts were made in an attempt to combine
various observed components into a coherent model to understand star-formation
activity in the complex (Heyer et al. 1989; Carpenter 1995a,b; Chavarr\'{\i}a et al. 2008).
Here, we extend this work with our new observations, as part of our ongoing
mutiwavelength investigation of star formation in and around Galactic \hii regions
(e.g. Ojha et al. 2004a, 2004b, 2004c; Tej et al. 2006; Samal et al. 2007; 
Pandey et al. 2008; Samal et al. 2010). We studied the nature of stellar population 
in the gas ridge between S255-S257 with high resolution NIR observations 
covering an area of $\sim$ 4$^{\prime}$.9
$\times$ 4$^{\prime}$.9 of the gas ridge,
along with radio continuum imaging at 14.94 GHz.
Using spectroscopic observations we attempted to establish that the ionizing 
sources of the S255, S257 and the compact \hii regions of
S255-IR cluster share a common plane in the sky. In order
to unravel the star formation history, we also identified and characterized the 
stellar population beyond the gas ridge, using optical photometry in 
conjunction with mid-infrared (MIR) photometry
from Chavarr\'{\i}a et al. (2008), for an area of $\sim$ 13$^{\prime}$ $\times$ 13$^{\prime}$. 
We used Robitaille's SED models for subsets of stellar sources 
in the gas ridge as well as outside to assess their
likely evolutionary status. Our findings suggest that 
the sources in the gas ridge show signature
of younger population. Since the gas ridge is sandwiched between two optically visible \hii regions,
we tried to interpret a possible star formation scenario in the gas ridge.
We have organised this paper as follows: In Section 2,
we describe our observations and the reduction procedures.
Section 3 describes spectral type of the ionizing sources of optically visible \hii regions
and their possible associations. Section 4 describes the general morphology and nature of stellar
sources and embedded cluster S255-IR. In Section 5, we discuss the age and mass of the stellar
sources beyond the gas ridge. Section 6 describes the SED modeling of stellar sources.
Section 7 is devoted to discussion on star formation scenario in the complex.
We present the main conclusions of our results in Section 8.

\section{Observations}

\subsection{Near-Infrared}

The imaging observations of the S255 SFR 
centered on $\alpha_{2000} = 06^{h}12^{m}53^{s}$,
$\delta_{2000} = +17^{\circ}59^{\prime}21^{\prime\prime}$
in the $J$ ($\lambda$ = 1.25  $\mu$m), $H$ ($\lambda$ = 1.63
$\mu$m), and  \ks ($\lambda$ = 2.14  $\mu$m) bands were obtained on 2000
October 13 with the 2.2 m UH telescope and SIRIUS, a
three-color simultaneous camera (Nagashima et al. 1999; Nagayama et al. 2003).
In this set up, each pixel of the 1024 $\times$ 1024 HgCdTe array 
corresponds to 0$^{\prime\prime}$.28  
yielding a field of view (FOV) of 
$\sim$ 4$^{\prime}$.9 $\times$ 4$^{\prime}$.9  on the sky.
A sky field which is 10$^{\prime}$ south of the target 
position (centered on $\alpha_{2000} = 06^{h}12^{m}53^{s}$,
$\delta_{2000} = +18^{\circ}09^{\prime}26^{\prime\prime}$) was also observed 
and used for sky subtraction.
We obtained 18 dithered frames each of 20s, thus giving a total
integration time of 360s in each band.
The observing conditions were photometric and
the average FWHM during the observing period 
was $\sim$0$^{\prime\prime}$.7-0$^{\prime\prime}$.9.
Photometry was done using the PSF algorithm of DAOPHOT 
package (Stetson 1987) in IRAF\footnote{IRAF is
distributed by the National Optical Astronomy Observatories, USA}, 
following the same procedure described in Samal et al. (2010).
The photometric calibration was done by observing the standard star P9107 in the faint
NIR standard star catalog of Persson et al. (1998) at air masses
closest to the target observations.
A comparison of present photometry with that of 2MASS photometry
as a function of $J$ mag for the common sources  is shown in Figure 2, where
the 2MASS sources with photometric error $\leq$ 0.1 mag and `read-flag'
values between 1 to 3 are selected to ensure good 
quality (see Samal et al. 2010 for description).
We find that the photometric scatter in  magnitude increases at the
fainter end, whereas the mean photometric uncertainties
are $\le$ 0.1 mag in all the three bands. 
Absolute position calibration of the detected sources was achieved 
using the coordinates of 20 isolated bright 2MASS sources. 
The astrometric accuracy is estimated to be $\sim$ 0$^{\prime\prime}$.05.
To plot the data in color-color (CC) and color-magnitude (CM) 
diagrams, the $JHK_{\rm s}$ data were transformed to California 
Institute of Technology (CIT) system using the relations
given by Nagashima et al. (2003).

\subsection {Optical}

\subsubsection {Photometric Observations}

The CCD $BV{(I)}_c$  observations of the S255 region centered on
$\alpha_{2000} = 06^{h}12^{m}54^{s}$,
$\delta_{2000} = +17^{\circ}59^{\prime}11^{\prime\prime}$, 
were acquired on  2004 December 15,
using the CCD camera mounted on  the 104-cm Sampurnanand telescope 
(ST) of ARIES, Nainital (Sagar et al. 1999). In this set up,
each pixel of the CCD corresponds to 0$^{\prime\prime}$.37 and 
yielding a field of view of
$\sim$ 13$^{\prime}$ $\times$ 13$^{\prime}$  on the sky. 
The FWHM of the star images was $\sim$ 2$^{\prime\prime}$.5.
To determine the atmospheric extinctions and to calibrate the CCD systems,
we observed standard stars in the SA98 field (Landolt 1992) and short exposures of the
target field using the ST on 2009 October 14.
The observing conditions were photometric and
the average FWHM during the observing period was $\sim$ 1$^{\prime\prime}$.8.
The CCD data frames were reduced using DAOPHOT-II software 
(Stetson 1987) following the same procedure described in Samal et al. (2010).
The standard deviations in the standardization residual, $\Delta$,
between standard and transformed $V$ magnitude and colors are  found to be less than 0.02 mag.
We generated secondary standards from the short exposure images on 2009 October 14 to standardize
the data observed on 2004 December 15. 
The mean and standard deviation of the photometric difference
between the common stars of the target sources at $B$, $V$ and $I_c$ bands on both the nights
are 0.009 $\pm$ 0.037, 0.006 $\pm$
0.031 and 0.016 $\pm$ 0.054 respectively, therefore, we expect a calibration error less than $\sim$ 0.05 mag
in all the bands. When brighter stars were saturated on deep exposure frames, their magnitudes have been
taken from short exposure frames to make the final catalog. Since the photometric errors become large ($\ge$ 0.1 mag)
for stars fainter than $V\simeq21.6$ mag, the measurements beyond this magnitude limit are not used in the
present study. 
An astrometric solution for the detected stars in the frame was determined 
using 15 isolated moderately bright stars with their positions from  
the 2MASS catalog and a position accuracy 
of better than $\pm$0$^{\prime\prime}$.1 has been achieved.

\subsubsection {Spectroscopic Observations}

We obtained optical spectra of the ionizing sources of S255 and S257, 
using 2-m telescope at IUCAA Girawali Observatory (IGO) on 2008 January 11.  
We used IUCAA Faint Object Spectrograph (IFOSC) 
with a silt of 1$^{\prime\prime}$.5 width in combination with a grism (IFOSC7) 
in the wavelength range 3800-6840 \AA~  at a spectral dispersion of  1.4 \AA~ pixel$^{-1}$.
The spectra of the ionizing sources of S255-2a
and S255-2b were obtained using  2.01-m Himalayan {\it Chandra} Telescope
(HCT) on 2009 December 20 and 2010 January 18, respectively. The instrument
is equipped with a  SITe 2K $\times$ 4K pixel CCD. 
We used Himalaya Faint Object Spectrograph Camera (HFOSC) with a silt of 1$^{\prime\prime}$.9 width in
combination with a grism (Gr 7) in the wavelength range of 3500-7000 \AA~ at
a spectral dispersion of 1.45 \AA~ pixel$^{-1}$.
 The observing conditions were photometric, with a typical seeing
of 1$^{\prime\prime}$.2 to 1$^{\prime\prime}$.3 during the nights.
The one-dimensional  spectra  were extracted and  wavelength calibrated
following the procedure described in Samal et al. (2010).  
An rms scatter of 1 \AA~ is expected in the wavelength calibration. 

The log of observations is given in Table 1.

\subsection {Radio Continuum Imaging}

We have also analysed the data of S255-IR at
 14.94 GHz, taken from the National Radio Astronomy Observatory (NRAO) 
data archive
(Project ID: AM697), obtained using the Very Large Array (VLA)
in the D configuration. The observation was carried out in
28 November 2001. The flux calibrator
was 0542+498 and 0625+146 was used as a phase
calibrator. The NRAO AIPS was used for the reduction of the data.
The estimated uncertainty of the flux calibration was within 5\%.
The image of the field was formed  by the Fourier inversion and cleaning
algorithm task IMAGR, with a Briggs weighting function (robust factor = 0)
halfway between uniform and natural weighting. We applied two iterations of
phase self-calibration to improve the map. The resulting image has a beam
size of $\sim$ 4$^{\prime\prime}$.4 $\times$ 4$^{\prime\prime}$.2 and the rms noise in the map
is 0.11 mJy/beam.

\section {Ionizing Sources and the Ionized Gas Distribution}

The one-dimensional spectra of the ionizing sources of S257, S255, S255-2a, and S255-2b
in  the  range  3950-4950 \AA~are shown in Figure 3. We classified the observed spectra by
comparing with those given by Walborn \& Fitzpatrick (1990).
In the case of O-type stars, the classifications of different subclasses are based on
the strengths of the \heii line  at   4541 \AA~ and \hei line 
at  4471 \AA. The equal strength of the lines
implies a spectral type of O7, whereas the lesser the strength of  4541 \AA~
line as compared to  4471 \AA~ line implies a spectral type
later than O7. The O-type luminosity class V spectra also have a 
strong \heii 4686 \AA~  
in absorption. The main  characteristics of early B-type stars are the absence of
\heii lines, which are still present in late O-type stars. If
the spectrum displays \heii line at  $\sim$ 4200 \AA, and
weak \heii line at 4686 \AA, along with OII/CIII blend at
$\approx$ 4650 \AA, this indicates the spectral type of the star
must be earlier than B1. Specifically, \heii line at  4200 \AA~ is seen in stars
of spectral type B0.5-B0.7. The presence of the \hei lines in absorption
constrains the spectral types to earlier than B5-B7. The \hei lines may however
be affected by the intense nebular emissions, which may fill the stellar
 absorption features of Balmer lines. It is not easy to subtract such
emissions, which may be misleading to the spectral classification.
The  absence of significant \siii line at 4128 \AA~ and
presence of very weak \mgii 4481 \AA~ line in the spectra indicate a 
spectral type earlier than B2 and the
contributions of these lines increase towards spectral types later than 
B2 (Walborn \& Fitzpatrick 1990).
By comparing the strength of various lines, we classified the most 
probable spectral types of the ionizing sources 
of S255, S257 and S255-2a as O9.5, B0.5 and B2 respectively.
The poor signal-to-noise ratio
of the spectrum of the ionizing source of S255-2b limits the accuracy of classification,
but the spectrum
shows the signature of an early-B type star, possibly B2-B3.
The luminosity criterion depends mainly upon
the ratios of \Siii 4553 \AA~to \hei 4387 \AA~and \siiv 4089 \AA~to \hei 4121 \AA, 
which shows a smooth progression along the main-sequence (MS) points to supergiant class.
Though the assessment of luminosity class is difficult from the 
present spectra, we favor
the luminosity as class V for the ionizing sources of S255 and S257
due to the presence of \heii 
4686 \AA~and \Siii 4553 \AA~lines in absorption. As both S255-2a and S255-2b have  characteristics comparable to those of compact \hii regions (Snell \& Bally 1986),
they are therefore not dynamically evolved objects. According to Meynet et al. (1994),
an 18 $M_\odot$ star (corresponds to B0 spectral type) has MS lifetime of $\sim$ 8 Myr. 
Therefore, the luminosity class V seems to be more plausible for 
the ionizing sources of S255-2a and S255-2b.

To obtain the  distance of S255-2a, we estimated the visual extinction ($A_V$) 
as 5.30 mag of the ionizing source, using a standard value of R$_V$ $=$ 3.12 
and adopting an intrinsic color  
$(B-V)_0$ = -0.24 (Cox 2000) for a B2 MS star, along with our measured photometric 
color $(B-V)$ = 1.46. Using the M$_V$ -spectral type calibration table 
from Russeil (2003), the distance turns out to be 2.6 kpc. For the ionizing 
sources of S255 and S257 regions, we adopted magnitudes and colors from 
Moffat et al. (1979) as the sources are saturated in our photometry.
For S255, we find $E(B-V)$ = 1.18 mag using $(B-V)$ = 0.88
and $(B-V)_0$ = -0.30, which corresponds to $A_V$ = 3.68 mag. 
Similarly, for S257, we find $E(B-V)$ = 0.87 mag using $(B-V)$ = 0.57 
and $(B-V)_0$ = -0.30, which corresponds to $A_V$ = 2.71 mag. Following
the same procedure as discussed above, the distance turns out to be 2.7 and 2.3 kpc 
for S255 and S257, respectively. The above analysis suggests 
probable association among the sources.
The resulting average distance (2.53 $\pm$ 0.21 kpc) for the complex agrees 
well with the average distance 2.46 $\pm$ 0.17 kpc obtained  by Russeil, Adami, \& Georgelin (2007). There is however some ambiguity concerning the distance to the region. 
A number of authors suggested that the whole region corresponds to a cloud complex at
2.3 - 2.6 kpc (Chavarr\'{\i}a et al. 2008 and references therein), which is
also supported by our distance estimation.
Therefore, in the present work we have adopted a distance
of 2.5 kpc for further analysis. However, it is worth mentioning a recent
study by Rygl et al. (2010) of the S255-IR region,
which reported a distance of $\sim$ 1.6 kpc.

The S255-IR region is a cluster forming environment which contains a number of
OB-stars (Howard, Pipher, \& Forrest, 1997), of which 
only two are visible in the optical band.
Previous observations at 5 GHz show that this SFR usually contains four 
centimeter continuum sources (Snell \& Bally 1986).
Since the emissions from \uchii~and hyper-compact \hii regions are optically thick at
this frequency, we analysed the radio continuum emission at 14.94 GHz obtained with VLA. 
Figure 4
shows contours of the  14.94 GHz radio free-free emission overlaid onto the  \ksb image. 
The map mainly shows five compact radio emissions and four of them are 
marked in the figure as per nomenclature by Snell \& Bally (1986).
However, we detected one extra source that has not been previously identified in the
literature. Though the 5 GHz map by Snell \& Bally (1986) shows some weak radio emission
approximately at the position of the source, the flux value of the source is not mentioned
in their paper. Hence, to conclude whether the source is an
extragalactic radio object or the true \uchii~region, one needs spectral
index measurement which requires emission at two frequencies.
Due to unavailability of the flux
measurement at other frequencies, in the present analysis, we tentatively do not consider
this radio source as a possible \uchii~region. However, the spectral indices 
between 5 GHz and the 14.94 GHz of the other radio sources suggest 
thermal origin. The regions of recent star formation in the \ksb image are traced by the
bright nebulosity, whereas the \hii regions detected in centimeter surveys trace 
the high mass star-forming environment. It is well established that formation 
of high mass stars in young complexes is accompanied by a cluster of less massive sources. 
Hence, the NIR point sources in Figure 4 are likely candidates of intermediate to
low mass stars. Out of the five compact radio emissions, the most intense emission
associated with components S255N and S255-2c seems to be elongated along the
east-west direction. 

To derive the physical properties of the thermal radio sources, we measured 
peak position, 
integrated flux density and angular diameter deconvolved from the beam by fitting
a two-dimensional elliptical Gaussian using the task JMFIT in AIPS.
We calculated the physical parameters assuming the simplest geometry of 
uniform spherical model. The physical parameters are: $n_e$, the rms electron 
density; $EM$, the emission measure; $\tau$$_c$, the optical depth; 
$N_{Lyc}$, the number of Lyman continuum photons for a
spherically symmetric \hii region and the ZAMS spectral type of 
the ionizing source. The ZAMS spectral type has been estimated from the
$N_{Lyc}$ and the calibration table of Panagia (1973), for an
electron temperature of 10$^{4}$ K. These parameters were derived from 
the measured angular diameter and integrated flux density and 
using the formulae given by Panagia \& Walmsley (1978) and 
Mezger \& Henderson (1967) for spherical, homogeneous nebulae.
The derived physical parameters are listed in Table 3. Since we derived 
the average properties, the rms electron density of the \hii regions
is expected to be lower than the density at the peak positions. 
The spectral types suggest that all the sources are ionized by 
massive early B-type ZAMS stars and their physical parameters point to the
evolutionary class between compact and \uchii~regions (Kurtz et al. 2000). 
Since the compact and \uchii~regions are not dynamically evolved objects, 
this indicates that the massive star formation has started recently at 
various locations of the gas ridge, whereas, the larger sizes and 
lower electron density values estimated from the 1.4 GHz observation 
by Snell \& Bally  (1986) for the \hii regions S255
(size $\sim$ 1.16 pc, n$_e$ $\sim$ 100 cm$^{-3}$) and 
S257 (size $\sim$ 0.87 pc, n$_e$ $\sim$ 111 cm$^{-3}$), indicate that 
both are in an advanced stage of evolution. 
This confirms that the ionizing sources of S255 and S257 are in
a more evolved stage than the massive stars in the gas ridge.

\section{Embedded Cluster}

Figure 5 represents a false $JHK_{\rm s}$ color-composite image
($J$, blue;  $H$, green; and $K_{\rm s}$, red) of our observed region
centered on the cluster S255-IR. In Figure 5, a clustering of stellar 
sources is apparent at the center of the image. The figure also 
displays several red sources, indicating the presence of likely young 
stellar population still deeply embedded in
the molecular cloud, whereas, bluest stars mainly seen towards the 
eastern and western parts of the image, are likely evolved/foreground 
objects of the field. A prominent $K_{\rm s}$-band nebulosity is seen 
at the center of the cluster, with a dark lane surrounding it.
The positions of the ionizing sources are marked with
circles in the enlarged version of the color-composite image shown in the right
panel of Figure 5, which reveals that most of the red sources in
the vicinity of S255-IR lie towards the western side, possibly indicating
that more young stars are concentrated in this direction.
Apart from the main clustering at the center, patchy nebulosity and
diffuse wisp emissions are also seen in the image and many 
red sources appear to be projected on these emissions. The 
single dish observations have established the presence 
of a large column of gas,
elongated along the lane between  S255 and S257 (Klein et al. 2005). 
Hence, these red sources are possibly young protostars that  
are associated with  the gas reservoir.

\subsection{Nature of the Stellar Sources}

We now discuss the nature of the associated low mass stellar population using 
our NIR observations.
To examine the nature of 
the stellar population in the gas ridge, we used NIR CM ($H-K$/$K$)
and CC ($H-K$/$J-H$) diagrams of the sources detected in 
$HK_{\rm s}$ and $JHK_{\rm s}$ bands, 
respectively, by comparing with the CM and CC diagrams
of a nearby control field. Since the extinction is more 
sensitive to the shorter wavelength ranges and also young 
protostars are more prominent at longer wavelengths, we use the sources
detected only in $H$ and $K_{\rm s}$ bands to construct $H-K/K$ CM diagrams.
Figure 6 shows the CM diagrams for the S255-IR SFR and for the
nearby control field, where the nearly vertical solid lines 
represent ZAMS at a distance of 2.5 kpc, 
reddened by $A_V$ = 0, 2.5, 20, 40 and  60 mag, respectively.
The parallel slanting lines represent the reddening vectors to
the corresponding spectral type. We would like to mention that
the uncertainty in spectral type determination is mainly due to the error in the
distance determination as well as the \ksb excess of the individual sources.
A comparison with the control field reveals that the majority of the
cluster population shows an apparent separation from the foreground field 
stars distributed at ($H-K$) $\sim$ 0.2 mag. Figure 6 manifests that 
the majority of the sources in the S255-IR region are likely to be 
reddened MS members and/or pre-main-sequence (PMS) sources with 
strong infrared excess emission. 
Most of the sources with $H-K$ $\geq$ 2, not detected in the $J$-band, 
are likely candidates for young sources with disks and envelopes.
Massive stars ($\ge$ 8 $M_\odot$) do not have PMS phase and 
appear on the ZAMS while still embedded in the molecular cloud, 
whereas, lower mass stars spend a significant amount of time in the PMS 
phase. Therefore, in Figure $6a$, all the  sources
may not be reddened ZAMS stars. There are 22 sources 
in Figure 6 which appear to be massive with spectral type earlier than B3.
Probably, these sources appear luminous due to
excess emissions in the \ksb because one would expect radio free-free 
emissions in the vicinity of such sources.
However, we detected only four compact radio emissions at 14.94 GHz in the vicinity of
SFRs S255-IR and S255-N. The map is sensitive down to 0.11 mJy/beam at 14.94 GHz,
which corresponds to $\sim$ 2.9 $\times$ 10$^{46}$ number of Lyman continuum
photons per second for an optically thin thermal source situated at a distance of 2.5 kpc, for an electron
temperature  of 10$^4$ $K$. 
This corresponds to a ZAMS spectral type  close to B2 (Panagia 1973) and
therefore sets an upper limit of $\sim$ 10 $M_\odot$, for the sources that do not show compact
radio emissions.
The sources associated with the radio emissions are  
marked in the $H-K/K$ CM diagram (Figure 6), following the same nomenclature by Howard et al. (1997).
Source IRS3 corresponds to the radio peak S255-2a, the strongest radio continuum source and
the bluest of all the stars of the S255-IR cluster. The CM diagram indicates the
star as B2 type. The source IRS2 corresponds to the  radio peak S255-2b 
and is situated $\sim$ 2$^{\prime\prime}$ southwest of S255-2a. The CM 
diagram suggests the source as B2-B3 type. The infrared spectral type 
estimation for both the sources is consistent with the radio estimation at 14.94 
GHz within a subclass. The position  of star IRS1b coincides with the radio
source S255-2c and the spectral type of the star appears to be earlier than O5,
whereas the radio estimation suggests a spectral type of B1. Since the star shows large $\sim$ $H-K$
color and is not detected in the $J$-band, it is probably 
the youngest among the three radio sources.
The polarization maps by Tamura et al. (1991) at $K$-band show a 
presence of a dusty disk towards this star.
Therefore, we believe that 
the discrepancy of infrared and radio luminosities is caused by the presence 
of excess emission in \ksb due to the circumstellar disk.
We do not discard the possibility of a few reddened background field stars, which might
have contaminated the CM diagram. However, by taking into account the distribution of
the high column density of \tco molecular gas (Chavarr\'{\i}a et al. 2008), it seems
unlikely that the stellar sources in the gas ridge consist of significant background
contamination.
         
To characterize the different populations of the region,
we use the $H-K/J-H$ CC diagram, as shown in Figure 7. The solid thin  and thick dashed
curves, taken from Bessell \&  Brett (1988), represent the loci  of early MS and
late giant  stars, respectively. The left and middle parallel dashed lines are the 
reddening vectors for early MS and late giant stars (drawn from the base
and tip of the two branches), respectively. The dotted line represents  the locus of
classical T Tauri stars (CTTS) (Meyer, Calvet, \& Hillenbrand 1997) and the right parallel dashed line represents the
reddening vector drawn from the tip of the locus of CTTS. 
The reddening  vectors are plotted
using $A_J/A_V$ = 0.265, $A_H/A_V$ = 0.155, $A_K/A_V$  = 0.090 for the 
CIT system (Cohen et al. 1981). The sources between the left and middle reddening vectors
are considered to be either field stars or reddened MS stars or weak-line T Tauri stars, 
whereas sources with NIR-excess usually fall to the right of the 
middle reddening vector and 
are considered as YSOs with disk and envelope (Lada \& Adams 1992).
The NIR CC diagram can also be
contaminated by foreground, background field stars and evolved stars. Therefore, to distinguish
the cluster members and YSOs from these objects, we used a statistical approach by comparing the
distribution of the sources in the NIR CC diagram of 
a nearby control region as shown in Figure 7b.
Comparison reveals that sources lying between the left and middle reddening vectors 
with ${\it (J-H)}$ $\geq$ 1.1 and the sources to the right of the middle reddening vectors 
are likely to be members and YSOs of the region. We do not rule out
the possibility that a few massive and 
intermediate mass members  may fall below with ${\it (J-H)}$ $\leq$ 1.1,
if they are reddened by the same amount like IRS3 and IRS2. However, considering the
initial mass function (IMF) is dominated by the lower mass stars, 
most of the members should have ${\it(J-H)}$ $\geq$ 1.1
and are possibly reddened MS members or PMS stars with inner circumstellar disk.
The evolutionary stages of these sources are difficult to estimate 
until we know about their SEDs. As can be seen 
in the CC diagram of the control field,
the NIR-excess zone is free of field populations, therefore  83 sources 
found in the NIR-excess zone in Figure 7$a$ are likely to
be YSO candidates. However, this number should be considered 
as a conservative lower limit,
as there are also many sources detected only in $H$ $\&$ $K_{\rm s}$ bands and 
only in $K_{\rm s}$ band, but not in the $J$ band.
The fraction of YSOs increases with high sensitivity observations 
at $L$- and $M$-band (e.g., Hoffmeister et al. 2008). 
{\it Spitzer}-IRAC observations by Chavarr\'{\i}a et al. (2008) uncover
several new clusters/groups as well as distributed population
of YSOs in the direction of S255-S257 complex. Giving emphasis to
the distribution of IRAC classified Class II and Class I sources,
they found the ratio of the Class I/Class II sources is highest
along the direction of the gas ridge between S255 and S257. 
In Figure 7$a$, squares represent the distribution of 21 
IRAC classified Class I and Class II YSOs and crosses denote the 115 
IR-excess YSOs by Chavarr\'{\i}a et al. (2008). It is important to  
note that many sources appear as 
normal stars in the NIR CC diagram, whereas they 
show IR-excess in the {\it Spitzer} bands.
This gives an important clue that many sources that appear
likely to be reddened MS stars in the NIR CC diagram,
could in reality be PMS sources with disks.  
In order to discuss the  star-formation activity in the gas ridge, 
we used the spatial distribution of the 83 YSOs identified from 
the NIR CC diagram and 26 sources having $H-K$ $>$ 2, along with
the YSO candidates identified by Chavarr\'{\i}a et al. (2008) 
(see Section 4.4).

\subsection {The \ksb Luminosity Function}

The \ksb  luminosity function (KLF) of an embedded
cluster is useful in constraining the age of the cluster (Megeath
et al. 1996; Ojha et al. 2004b,c). In order to derive the observed KLF, one
needs to apply corrections for (1) the incompleteness of the star
counts as a function of the $K_{\rm s}$ magnitude, and (2) the field star
contribution in the line of sight of the cluster.
The completeness was determined through artificial star 
experiments (see Ojha et al. 2004c). To  correct for the 
foreground and background contamination, we used the stellar population 
synthesis   Galactic model (Robin et al. 2003),
using a similar procedure described in Ojha et al. (2004c). 
The star counts were predicted in the direction of S255-IR 
using the average extinction 
to the embedded cluster \av~= 10 mag for a distance of 2.5 kpc.
After the determination of  fraction of the contaminating stars over the 
total model counts, we  scaled the model prediction to the star counts in the 
reference field, and subtracted the foreground and background
contamination from the KLF of the S255-IR region.

After correcting for the foreground and background star
contamination and photometric completeness, the resulting KLF
for the S255-IR region is presented in Figure 8.
In Figure 8, a power law with a slope,
$\alpha$ [$d N(m_K)/dm_K \propto 10^{\alpha m_K}$, where $N(m_K)$
is the number of stars brighter than $m_K$] has been fitted to the KLF using a 
linear least-squares fitting routine. The KLF of the S255-IR region shows
a slope of $\alpha$  = 0.17 $\pm$ 0.03 over the magnitude range
12.5 - 17 in the $K_{\rm s}$-band (solid line in Figure 8). 
The derived KLF slope is lower than those
generally reported for the young embedded clusters ($\alpha \sim  0.4$,
e.g., Lada et al. 1991; Lada, Young, \& Greene 1993; Lada \& Lada 2003), 
indicating a younger age of $\sim$ 1 Myr.
However, a break in the power law can be noticed at $K_{\rm s}$ = 15.25
mag and the KLF seems to be flat in the magnitude range of 15.25 - 17.75.
The slope of the KLF in the magnitude range of 12.5 - 15.25 (dashed line in 
Figure 8) comes out to
be 0.25 $\pm$ 0.05 which is still lower than the average value of slopes
($\alpha \sim  0.4$) for young clusters of similar ages
(Lada et al. 1991; Lada, Young, \& Greene 1993; Lada \& Lada 2003). 
A turn-off in the KLF has
also been observed in a few young clusters, e.g. at $K$ $\sim$ 14.5 mag,
$K$ $\sim$ 16.0 mag and $K$ $\sim$ 15.75 mag in the case of Tr 14
(distance $\sim$ 2.5 kpc; Sanchawala et al. 2007), NGC 7538
(distance $\sim$ 2.8 kpc; Ojha et al. 2004c) and NGC 1624
(distance $\sim$ 6.0 kpc; Jose et al. 2011), respectively.

\subsection {Mass Spectrum of Stellar Sources}

As discussed in Section 4.1, the excess emission in the \ksb may make the
YSOs luminous, hence the mass estimation on the basis of $H-K/K$ CM diagram 
will lead to an overestimation of the mass. To minimize the effect of NIR excess 
emission on mass estimation, we preferred $J$-band luminosity rather 
than $H$ or $K$ band. Figure 9 represents $J-H/J$ CM diagram 
for the sources detected in our observed area, where asterisks 
represent  YSOs selected from the NIR $H-K$/$J-H$ CC diagram. We 
assumed an age of 1 Myr to estimate the mass, 
which would be appropriate 
because the mean age of 16 YSOs in the direction of the embedded cluster 
is found to be $\sim$ 1.2 Myr (see Section 6).
The solid and dashed curves in the figure denote the loci of 1 Myr
PMS isochrones by Siess et al. (2000) for 
1.2 $M_\odot$ $\leq$ $M$ $\leq$  7 $M_\odot$ and Baraffe et al. (1998) for
0.05 $M_\odot$ $\leq$ $M$ $\leq$ 1.4 $M_\odot$, respectively. The 
dotted slanting
lines are the reddening vectors for  4, 0.5, 0.2 and 0.1 $M_\odot$ stars for 
1 Myr isochrones, respectively.
For an assumed age of 1 Myr, the lowest mass detected in 
our $J$-band image is about 0.1 $M_\odot$ for \av = 5 mag, 
and is about 0.3 $M_\odot$ for \av = 11.5 mag. In Figure 9,  
squares represent the {\it Spitzer}-IRAC identified Class I and Class II 
YSOs, which indicates that the IRAC observations
in the gas ridge are sensitive only down to 0.5 $M_\odot$. Whereas the 
IR-excess YSOs (crosses) identified
using NIR plus IRAC bands are sensitive down to 0.2 $M_\odot$ for \av = 5 mag.
The lower sensitivity of IRAC observations are most likely due to the combined effect of 
saturation, bright polycyclic aromatic hydrocarbon (PAH) emission towards 
S255-IR region and lower 
resolution of the IRAC data. 
Indeed Chavarr\'{\i}a  et al. (2008) found that the area with PAH emissions 
in the complex decreases the sensitivity to detect faint sources at 5.8 and 8.0 $\mu$m compared 
to the area free of PAH emissions. These are the possible causes which might 
have limited the 
sensitivity of the IRAC bands to detect YSOs upto 0.5 $M_\odot$, less sensitive than 
our NIR detection ($\sim$ 0.1 $M_\odot$). 
The reddening vectors for the 
assumed age suggest that most of the YSOs have masses in the range of 0.1 - 4 $M_\odot$.
With the assumed age, the distribution of YSOs with wide colors (Figure 9)
probably indicates the combined effect of variable extinction, weak
contribution of excess emission in $J$ and $H$ bands and/or sources in
different evolutionary stages (see also Wang et al. 2011). 
We note that the estimated stellar masses
from the infrared CM diagrams rely on the uncertain
age and distance. The ambiguity is more severe among the early
B-type stars because the inferred stellar mass for a 4 $M_\odot$ star from the 1 Myr PMS isochrone
can be as high as 18 $M_\odot$ when estimated from the MS isochrones of 
a younger age. For this purpose,
we have also drawn the reddening vectors (slanting dashed-dotted lines) 
from the tip of B0 and B5 ZAMS (dashed-dotted vertical curve) locus. 
The concept of PMS is not applicable to 
stars above 8 $M_\odot$ since the birth line and the ZAMS unify at those mass 
levels (Palla \& Stahler 1993). Therefore, the ionizing sources IRS3 and IRS2,
are likely to be reddened ZAMS massive stars.
It is worth pointing out that Howard, Pipher, \& Forrest (1997), 
based on the stellar fit of the
observed NIR data points, indicated that the region is rich in B-type stars,
containing 9 B0-type ($\sim$ 18 $M_\odot$) and 19 B5-type ($\sim$ 6 $M_\odot$) MS
stars within an area of 1.69 arcmin$^2$ centered on 
S255-IR cluster ($\alpha_{2000} = 06^{h}12^{m}53^{s}$,
$\delta_{2000} = +17^{\circ}59^{\prime}23^{\prime\prime}$).
The $J-H/J$ CM diagram does not reveal such massive members
(at least earlier than B0) for an assumed age of 1 Myr, as most of the stars have masses less
than 4 $M_\odot$. Moreover, one would expect radio emission from such sources.
The circumstellar disks are rare among the intermediate mass to massive PMS stars
with ages $\sim$ 1 Myr, therefore the mass uncertainty due to the circumstellar
disk should not be the major cause for massive members. The discrepancy
could be due to the contribution from the bright nebulosity to the photometric
magnitudes.

\subsection {Spatial Distribution of YSOs}

The infrared excess in the case of YSOs can be due to circumstellar disk/envelope
or weaker contribution from the reflected stellar radiation
of the dust emission. In any case, infrared excess represents the association
of young sources in a SFR and possibly traces the star formation history of a cloud.
Figure 10 shows the spatial distribution of YSOs from NIR CC diagram (asterisks), possible YSOs
detected only in the $H$ and $K_{\rm s}$ bands with $H-K$ $\geq$ 2 (filled circles) and
sources detected only in the \ksb (crosses). The  outer boundaries of 870 
$\mu$m dust continuum emission (Klein et al. 2005)
observed with a resolution of $\sim$ 26$^{\prime\prime}$ are shown with white 
dashed lines in the figure. This has the shape of
a bar extending in the north-south direction and separates 
the two evolved \hii regions S255 and S257. The outer boundaries 
of S255 and S257 are also marked with the red thick dashed lines,
which are extracted from the ionized gas traced by the radio continuum 
emission at 610 MHz observed with Giant Metrewave Radio Telescope (GMRT)
at a resolution of $\sim$ 5$^{\prime\prime}$.5 $\times$ 6$^{\prime\prime}$.5,
taken from Zinchenko et al. (in preperation). The yellow circles denote
 the positions of the three prominent 1.2 mm dust continuum clumps MM1, MM2 and MM3, revealed
by Minier et al. (2005) at a resolution of 24$^{\prime\prime}$.
The large crosses represent the positions of the radio continuum sources, three of which are found
in the vicinity of the S255-IR cluster and one is associated with the  S255N SFR. Most of the NIR YSOs are distributed in the close proximity  of the 
S255-IR cluster and are also projected in the direction of the 
ionized emissions of the evolved \hii regions. The majority of the 
YSOs with $H-K$ $\geq$ 2 and  those detected only in the \ksb  
are distributed along the bar of dust emission.
The sources detected only in the \ksb show a scattered distribution 
along the gas ridge, with an enhanced concentration 
towards the western side of the S255-IR cluster and in
the close vicinity of the molecular clump MM2. Towards S255N, MM1 and MM3, 
we do not see enhanced concentration of YSOs as seen towards S255-IR, possibly they
are in a more embedded phase of star formation.
In our observed NIR field, we identified  109 YSO candidates, which include 83 
from NIR CC diagram and 26 sources with H-K $>$ 2. Out of 109 sources, we find that 
only 40 in our observed region are common to the sources classified as YSOs by 
Chavarr\'{\i}a et al. (2008). Therefore, in  the present work we identified
69 new YSO candidates in the region.
To see the spatial distribution of all the YSO candidates, we overlaid the 
stellar surface number  density (SSND) contour
map on the \ksb image in Figure 10 using kernel  
method (Gomez et al. 1993; Silverman 1986). 
The SSND contour map was made using a total of 213 YSO
candidates, which include the 69 new YSOs detected by us and the 144 YSOs 
found by Chavarr\'{\i}a et al. (2008) 
using NIR and IRAC bands in the region observed by us. 
The stellar density distribution   is     smoothened     by     0.1 pc $\times$ 0.1 pc sampling box.
The contour levels are drawn from 20 stars/pc$^2$  with
an  increment of 30 stars/pc$^2$. The clustering towards S255-IR region is 
obvious in the SSND map along with substructures between S255-IR and MM1 and towards MM2. 
The main peak of the projected stellar density distribution was found to be $\alpha_{2000}=06^{h}12^{m}53^{s}$,
$\delta_{2000}=17^{\circ}59^{\prime}25^{\prime\prime}$, with central density of 180  stars/pc$^{2}$, 
whereas one can also see a second peak at $\alpha_{2000}=06^{h}12^{m}54^{s}$,
$\delta_{2000}=+18^{\circ}00^{\prime}06^{\prime\prime}$, with central density of 87  
stars/pc$^{2}$.  
This distribution confirms the presence of enhanced concentration of YSOs 
in the gas ridge compared to that towards S255 and S257 and is therefore 
a relatively more active site of star formation. 

\section{Stellar Population Beyond the Gas Ridge}

So far we have studied the properties of the stellar sources along or 
in the vicinity of the gas ridge. In order to correlate the star formation 
activity in the gas ridge with the stellar population  beyond the gas ridge, 
we used the optical CM ($B-V/V$ \& $V-I/V$) diagrams for the sources covering a field 
of $\sim$ 13$^{\prime}$ $\times$ 13$^{\prime}$ centered on S255-IR. 

In order to ascertain the age of the stellar members, we attempted quantitative 
age determination
with the help of the optical CM diagram, assuming that the emissions are 
purely photospheric and contribution from the disks and envelopes in 
the optical band is negligible. Figure 11 shows $B-V/V$ and
$V-I/V$ CM diagrams of all the observed stars. The distribution 
of the stars in the $B-V/V$ CM diagram resembles the foreground 
population distribution shown in the $H-K/K$ CM diagram of the 
control field (see Figure 6b), when constructed using 2MASS PSC. 
Hence, we presume that a majority of these stars are foreground stars. 
Moreover, the OB associations are generally not so conspicuous since they 
extend usually over a large area in the sky and most of the faint 
stars in the area are actually unrelated foreground or background stars.
The identification of the associated faint low-mass members among 
the field stars is rather difficult.
From the $K$-band extinction map, Chavarr\'{\i}a et al. (2008) have 
estimated a lower limit of the extinction value as 
A$_K$ = 0.4 mag (\av~$\sim$ 4.4 mag) for the members of the region. 
However, it is worth noting that the region shows variable 
extinction due to non-uniform distribution of the molecular material.
In Figure 11 we also plotted the ZAMS by Schmidt-Kaler (1982) 
using $E(B - V)$ = 1.41 mag (corresponds to \av~= 4.4 mag) and 
$E(V - I)$ = 1.25 $\times$ $E(B - V)$, for a distance of 2.5  kpc. 
It is worth pointing out that in comparison to $B-V/V$ CM diagram, 
$V-I/V$ CM diagram has a significant number of stars that  
fall to the right of reddened ZAMS. These could be 
low mass Galactic field populations and/or PMS stars 
of the region. The dynamical ages of the \hii regions S255 and S257 are
less than 2 Myr (Chavarr\'{\i}a et al. 2008). If we assume that the 
dynamical ages represent the age of the luminous exciting sources then we would expect
a number of contracting low mass PMS stars, because the star formation is
a continuous process and the contraction time of the YSOs to reach the MS increases
with decrease in mass. For example, the high mass ($>$ 8 $M_\odot$) 
stars take $\sim$ few $\times$ 10$^{5}$ yr to reach the MS, whereas, the low mass YSOs
($\sim$ $1$ $M_\odot$) can take 30 Myr to reach the MS.
In the absence of proper motion studies and spectroscopic
information, the distinction of true low mass members from
the contaminating field stars projected along the line of sight is difficult. 
Therefore,
we used the confirmed Class II, Class I and IR-excess YSOs identified by 
Chavarr\'{\i}a et al. (2008).
The 51 YSOs having optical counterpart are plotted in the $V-I/V$ CM diagram 
with circles, triangles and crosses, respectively (see Figure 11).
We derived the extinction values of  these YSOs  by tracing them back  along the
reddening vector to the CTTS locus or its extension in the $H-K/J-H$ CC diagram. 
The mean extinction value for
the sources above CTTS locus was found to be 4.6 $\pm$ 1.8 mag 
after excluding those sources that appear below the CTTS locus.
These sources are probably reddened Herbig Ae/Be type stars of the region. 
It is worth noting that the CTTS locus has its own intrinsic broadening due to colors 
of T-Tauri stars, which could also be a possible reason for the sources that fall below
the mean locus. We would like to mention that the approach to estimate the
extinction is an approximate one. However, the mean value of 
\av~($\sim$ 4.6 $\pm$ 1.8 mag) is in agreement with the
lower limit of \av~($\sim$ 4.4 mag) suggested by  Chavarr\'{\i}a et al. (2008).

Figure 12 shows the distribution of YSOs in the $V-I/V$ CM diagram, where
the Class II, Class I and IR-excess sources are shown with circles, triangles 
and crosses, respectively.
 To estimate the age of the stellar sources, 
the isochrone for 2 Myr for solar metallicity by Girardi et al. (2002) 
and PMS isochrones by Siess et al. (2000) for ages of 0.1, 0.5, 1, 3 and 4 Myr
have also been plotted. All the isochrones are corrected for a distance of 
2.5 kpc and visual extinction of 4.6 mag. The distribution of the PMS stars 
in the CM diagram implies a span in ages up to 4 Myr.
There are a few PMS stars located to the right of 0.1 Myr, these could be 
highly extincted sources. The age spread could also be an artifact of
the combination effects, e.g., variable extinction along the 
line of sight and variability. However, a similar age
spread has been observed in other SFRs as well 
(e.g., Jose et al. 2008; Sharma et al. 2007). A spectroscopic analysis 
may lead to a proper estimation of extinction, 
and a better picture of the age
of the stellar sources. To see the mass distribution of PMS stars, we use the grids of evolutionary tracks
by Siess et al. (2000), for stellar masses between 0.6 - 3.0 $M_\odot$, which are shown
with dashed-dotted curved lines in Figure 12. 
The  CM diagram indicates that PMS stars are 
younger than 4 Myr, with masses between 0.6 to 3.0 $M_\odot$.
It is to be noted that {\it Spitzer} observations 
reveal that the complex contains several small groups of
stellar sources (Chavarr\'{\i}a et al. 2008), 
possibly at different evolutionary stages. Therefore,
the observed age spread may represent the mixed population of different ages associated with
the molecular cloud. The YSOs included in the optical CM diagram (Fig. 12), 
are distributed
outside of the gas ridge between the two \hii regions S255 and S257, 
whereas, the YSOs detected in our
\jhk bands within the gas ridge are deeply embedded in the molecular 
cloud of \av~$\sim$ 10 mag, and are not detected in the optical bands.

\section {Physical Properties of the YSOs}

In the previous sections, we have identified YSOs at various locations and discussed
their nature with the help of the optical and NIR CM diagrams. In this section, we 
model the SEDs using the recently available
grid of models and fitting tools of Robitaille et al. (2006, 2007). The models are
computed using a Monte-Carlo based radiation transfer code (Whitney et al. 2003a,b)
using several combinations of central star, disk, in-falling envelope and bipolar cavity
for a reasonably large parameter space. The basic model consists
of a PMS star surrounded by a flared accretion disk and a
rotationally flattened envelope with cavities carved out by a
bipolar outflow. Interpreting SEDs using radiative transfer code is subject to 
degeneracies, and spatially resolved
multiwavelength observations can break the degeneracy. Therefore, we only fit the SEDs
to those {\it Spitzer}-IRAC YSO candidates, for which we have minimum of 6 data points 
in the wavelength range from 0.55 to 8 $\mu$m,
in order to constrain the parameters of stellar photosphere and circumstellar environment.
In this approach, we identified 31 sources detected in more than six 
bands in optical ($VI_c$), NIR ($JHK_{\rm s}$) and
IRAC (3.6, 4.5, 5.8 \& 8.0 $\mu$m) outside the gas ridge and 
21 sources having NIR $JHK_{\rm s}$ and 
IRAC counterparts in the gas ridge. Since the models are valid 
only for single isolated objects, 
out of 21 sources in the gas ridge, our \ksb image shows presence of 
multiple sources in 5 cases within the IRAC apertures 
used by Chavarr\'{\i}a et al. (2008) to extract 
the photometry. Therefore, we fitted SEDs to only 16 isolated sources
in the gas ridge. 
Since the \uchii~regions are not dynamically evolved objects, 
we also fitted SED to the ionizing source of S255-2c. The SED fitting tool fits each of the models to 
the data, allowing both the distance and external foreground extinction to be free parameters.
We give a distance range of 2.3 to 2.7 kpc (see Section 3) 
and $A_V$ is estimated using the NIR CC diagram. 
Considering the uncertainties that might have gone into the estimates,
we used the estimated value of $A_V$ $\pm$ 2 mag as the input parameter.
For the massive source associated S255-2c, we estimated $A_V$ 
from $H-K/K$ CM diagram.
We further set 10 to 15$\%$ error in NIR and MIR flux estimates due to
possible uncertainties in the calibration, extinction and intrinsic
object variability. As emphasized by Robitaille et al. (2007), 
their SED fitting tool does not claim to determine only the set of 
parameters that provide a good fit to the SED, but it helps to
find a range of reasonably well constrained parameters. 
Adopting a similar approach by Robitaille et al. (2007), 
we consider only those models to constrain the physical parameters, 
which satisfy the following equation :
\begin{equation} 
        \chi^2 - \chi^2_{\rm min} \leq 2N_{\rm data}, 
\end{equation}
where $\chi^2_{\rm min}$ is the goodness-of-fit parameter for the
best-fit model and $N_{\rm data}$ is the number of input observational
data points. Figure 13 shows examples of SEDs with the resulting models for
Class II, Class I and ionizing source of the \uchii~region S255-2c. In the
figure, a range of models are plotted in solid lines along with the best-fit
model. The values of the parameters for the YSOs outside and
inside the gas ridge are given in Tables 4 and 5.
In the tables, the listed parameters are:
the star  mass ($M_\star$), temperature ($T_\star$), 
stellar age ($t_\star$),  mass of the disk ($M_{\rm disk}$), disk accretion rate
($\dot{M}_{\rm disk}$), envelope accretion rate
($\dot{M}_{\rm env}$), foreground visual absorption ($A_V$) 
and the $\chi^2_{\rm min}$ of the best fit.
The tabulated parameters are obtained from the weighted mean and standard
deviation of all the  models that satisfy Eq. 1,  weighted by the inverse
square of ${\chi}^2$ of each model. The associated errors shown in 
Tables 4 and 5
are large because we are dealing with a large parameter space, with
limited number of observational data points. Additional observational data
points from  MIR to millimetre wavebands are required to constrain these parameters
more precisely.
Table 4 reveals that the age of the optically detected YSO candidates varies
between 0.5 to 5 Myr, with a mean age of $\sim$ 2.5 Myr, whereas the visual extinction varies
from 2.7 to 7.6 mag, with a mean around 4.4 mag. 
The mass of the YSOs shows a range from 0.9
to 3.0 $M_\odot$. These estimations are not distinctly different 
from the estimation based on the $V-I/V$ diagram.
Table 5 shows that the age of the majority of YSOs, 
along or in the vicinity of the gas ridge, 
varies from 0.1 to 2 Myr with a mean age of $\sim$ 1.2 Myr,
whereas the visual extinction varies form 4.5 to 22.1 mag. 
The derived cluster age ($\sim$ 2 Myr) is slightly lower than that obtained 
by Wang et al. (2011) from the intrinsic HR diagram. 
It is to be noted that the Wang et al. (2011) estimate was determined by  
adopting a distance of 1.6 kpc. Though direct comparison is not possible
because of different approaches to estimate age, it is worth to mention 
here that adopting a lower distance in the HR diagram makes the age older.
The mass of the YSOs shows a range from 0.6 to 7.6 $M_\odot$, implying 
we are sampling YSOs only upto 0.6 $M_\odot$.
For ionizing source of the \uchii~region S255-2c, the
acceptable SED fitting models suggest a massive star ($\sim$ 27 $M_\odot$), with an age of $\sim$ 0.1 Myr. 
The models also predict that the source is deeply embedded behind $\sim$ 42 mag of 
visual extinction and has an envelope accretion rate of 10$^{-3.4}$ yr$^{-1}$. 
Similarly, following the procedure given by 
 Chavarr\'{\i}a et al (2008), the dynamical age is estimated to be less than 10$^{5}$ yr for the compact and \uchii~regions, for an assumed 
density of 10$^{5}$ cm $^{-3}$. These estimations indicate that the massive stars  associated with
the gas ridge are younger in nature. It is also apparent from  Tables 4 and 5 that
the disk accretion rate of the YSOs outside the gas ridge is in
the range of 10$^{-7}$ $-$ 10$^{-8}$ $M_\odot$ yr$^{-1}$, less than that of YSOs in the gas ridge, which is in the
range of 10$^{-6}$ $-$ 10$^{-7}$ $M_\odot$ yr$^{-1}$. Similarly, the envelope infall rates for most of the 
YSOs inside the gas ridge are relatively higher than the YSOs outside the gas ridge.
These derived characteristics of the YSOs in the gas ridge are possibly 
the consequence of their younger age than the YSOs distributed outside. 
However, in the absence of MIR to millimeter data the above values should be 
treated with caution.

\section{Discussion on Star Formation Activity}

The derived characteristics of the YSOs along with the morphology of the complex 
and  physical properties of the associated gas and dust can be used to discuss the
evolutionary status of the star-forming sites and the overall star-formation 
activity of the complex.
 
\subsection{Early stage of star formation in the clustered environment}

An embedded cluster along with three prominent clumps MM1, MM2 and MM3 
(Minier et al. 2005) appear to be sandwiched between the
two evolved \hii regions S255 and S257. The clumps MM1 \& MM2 contain a similar amount of dense gas
of mass $>$ 200 $M_\odot$, and show a higher gas temperature of 40 K and are in
agreement with that of Zinchenko et al. (2009).
This is in contrast to MM3, which has a mass of $\sim$ 100 $M_\odot$ and the gas temperature is expected to be around
20 K. These observations led us to suggest that the temperature of MM1 and MM2 is
significantly higher than the
starless clumps, representing that the clumps are possibly internally heated by the 
protostars, whereas the clump
MM3 represents the  earlier stage of star formation as compared to MM1 and MM2.
It is to be noted that the clump MM2 lies close to the  infrared cluster S255-IR, where the cluster  harbors two
optically visible massive stars, along with several intermediate to low mass stars of different properties
detected in the \jhk bands, whereas, the sources in the vicinity of the clump MM2 are detected in the \ksb only,
and possibly represent a more embedded phase of star formation.
The clump MM2 is also associated with CH$_3$OH,  H$_2$O and OH masers
(Minier et al. 2005; Goddi et al. 2007).
CH$_3$OH maser traces earliest stages of massive star formation
(Walsh et al. 1998) and turns off shortly after the formation of
\uchii~region. The detection of maser emission towards the centre of
MM1, which has a characteristic lifetime of 2.5 $\times$ 10$^{4}$ to  4.5 $\times$ 10$^{4}$ yr
(van der Walt 2005), indicates this is not a dynamically evolved object.
It seems that the sources associated with the clump MM2 are most probably
younger than the sources associated with S255-IR. The sources associated
with the \hii regions S255-2a and S255-2b of the cluster S255-IR are not associated
with millimetre continuum emission and hence, presumably are in the later stage of evolution.
There was also evidence of bipolar outflows from the sources associated
with the S255-2c (Wang et al. 2011; Zinchenko et al., in preparation), along
with presence of dusty disk (Tamura  et al. 1991),
which supports the interpretation that the B-type stars may form via an accretion
outflow process similar to that of low mass stars. Both these phenomena indicate an
early evolutionary stage of the source (IRS1b) associated with
S255-2c, which is supported by
the dynamical age and the age derived from the SED fitting models (i.e age $\leq$ 0.1 Myr).
An interesting question is whether the sources in the vicinity of MM2 and
the massive stars of
S255-IR  have formed in the same region of space. Although presently we do not have
the necessary data to address this issue and reach a firm conclusion,
it is quite possible that both populations
are spatially unrelated. We also do not discard the possibility that they may represent different evolutionary stages of
the same cluster forming environment. For example, in the case of the high mass star-forming region, IRAS 20293+3952, Palau et al. (2007)
found that the region consists of starless cores, millimeter sources, \uchii~region and more evolved NIR sources.
They concluded that these sources are not forming simultaneously in this cluster environment. Rather, there
may be different generations of star formation and this could be a possible case for S255-IR. In several cases, as for example W3
(Feigelson \& Townsley  2008; Ojha et al. 2009) and IRAS 19343+2026 (Ojha et al. 2010), the massive embedded sources
are found to be associated with a cluster of  more evolved lower mass Class II and Class III sources, detectable in the NIR
bands. This suggests that, in these regions  the intermediate to massive protostars have formed after the first generation of
low-mass stars. The clump MM1 is associated with an \uchii~region identified in the 14.94 GHz map and is possibly ionized by a
B1 type star. Kurtz, Hofner, \& \'Alvarey (2004) detected a Class I CH$_3$OH maser at 44 GHz and an H$_2$O maser at 22 GHz in the
close vicinity of the \uchii~region.  The non-detection of ionizing source(s) of the associated \uchii~region in the
near to mid infrared band and the presence of  H$_2$  knots tracing an outflow (Miralles et al. 1997), indicate a very young age.
Since the initial phase of massive star formation is intimately linked to cluster formation and the fragmentation of the
molecular clouds, this indicate that S255N is a site of cluster formation in its early stage. Indeed,
Cyganowski, Brogan, \& Hunter (2007) found evidence of a massive protocluster with the help of high resolution
($\sim$ 4$^{\prime\prime}$.7 $\times$ 2$^{\prime\prime}$.7) interferometric observations at 1.3 mm continuum emission and detected three compact
cores with no infra-red counterparts. The masses of the cores range from
6 - 35 $M_\odot$. The identification of centimeter emission and multiple
millimeter continuum sources towards the S255N region indicates an ongoing (proto)cluster formation. 
The features discussed above support
the youthfulness of MM2 and MM1 regions and indicate the upper limit for star formation activity around MM2 and MM1
should not be more than $\sim$ 10$^5$ yr. The  clump MM3 shows no ionized gas emission in the proximity of the dust peak of
mass $\sim$ 100 $M_\odot$. For a typical star formation efficiency of 0.3 (c.f Lada \& Lada 2003), the clump would likely
form a star greater than 8 $M_\odot$. If this happens then it is a site for the earliest stages of star
formation in which the massive star has not yet reached the ZAMS. This indicates that it is likely to
be less than $\sim$ 10$^{5}$ years old because a single early B-star (M $\sim$ 10 $M_\odot$)
reaches the ZAMS in about 10$^{5}$ years.
It is worth mentioning that Wang et al. (2011) detected low velocity
outflows in MM3 (S255S) as compared to MM2 and MM1
and suggested that MM3 is likely in the collpase phase.
The lower temperature of the clump MM3
and non-detection of stellar sources in the \ksb  possibly indicates that MM3 is
in the earliest evolutionary stage.
The above discussion reveals that the star formation
is non-coeval along the gas ridge, where the stars of ages from
0.1 - 2 Myr are spatially distributed along the different locations of
the cloud.

\subsection {Surroundings of S255-S257 Complex}

The exciting stars of S255 and S257 are isolated 
sources, without any co--spatial clustering of low-mass stars around them
(Zinnecker et al. 1993). However, Chavarr\'{\i}a et al. (2008) hypothesize 
that a cluster (G192.63-0.00) is associated with S255. Future spectroscopic 
observations would confirm the association of G192.63-0.00 with S255.
Figure 14 shows the surroundings of the evolved \hii regions S255 and S257, 
where the white  background image represents the IRAC
8.0 $\mu$m  emissions which generally trace PAHs coming from  
the photodissociation regions (PDRs) at the surface
of the molecular cloud. Generally, these emissions are due to
the absorption of far-UV radiation at the surface of the molecular cloud, 
which  escape from the \hii region
and the PAHs within PDRs are excited and remit their energy at MIR wavelength. 
The green radio contours
show the ionized gas from radio continuum observations at 610 MHz 
(Zinchenko et al., in preparation). In radio, both S255 and
S257 appear almost circular in size. The figure also displays that 8 $\mu$m  
emissions are prominent at
the peripheries of S255 and S257, in the direction of the dark lane
and  are weaker towards the interiors of S255 and S257 from the lane, particularly for S257.
These emission features correspond to the dark patches seen in the optical band 
and also roughly coincide with CO emissions
found in these directions  by Chavarr\'{\i}a et al. (2008). 
The  $^{13}$CO  map by Chavarr\'{\i}a et al. (2008)
reveals that the distribution of the gas is rather non-uniform and clumpy, 
with its surface density reaching maximum along the
gas ridge, where new embedded star formation is expected to occur.

S255-S257 is a part of Gemini OB1 association. The star formation in OB association is often quite complex. One common property
of OB associations is that they are sub-structured and often consist of sub-groups of different ages.
Carpenter et al. (1995a,b) detected several lower luminosity sources and presumably low mass cores scattered
through the Gemini OB1 complex, whereas most of the massive dense cores are found adjacent to the optical \hii
regions and often show arc-shaped structure. They interpreted that the induced star formation in the massive dense 
cores is prominent in the complex due to sweeping and compression of the molecular material.
The complex S255-S257 shows a clear congregation of highly obscured
YSOs, Herbig-Haro objects, astronomical masers, compact 
and \uchii~regions. All these sources are concentrated preferentially along the 
vertical dense bar of gas and dust at the interface of the two \hii regions  
seen by the ionized  gas distribution at 610 MHz. The ionized gas emissions are delineated with PDR, and therefore indicate a definite
evidence of physical interaction of the \hii regions with the surrounding molecular material and is responsible for
the shaping of the ring like PDR structure.
The ionizing sources of the two adjacent  evolved \hii regions are
MS O9-B0 stars. If they have evolved in the same ambient density, then they might have similar ages.
The MS life time of  O9 star is of the order of $\sim$ 6.5 Myr.
It is difficult to constrain the precise age of the MS star with sufficient accuracy, to compare its
relative age with that of the age of YSOs detected in the gas ridge. 
However, the dynamical age of both the evolved \hii regions is
in the range of $\sim$ 0.8-1.6  Myr (Chavarr\'{\i}a et
al. 2008; Bieging et al. 2009), whereas, the low mass sources scattered in the complex show a  mean 
age of $\sim$ 2.5 Myr.
Low mass stars do not dramatically alter the environment in their vicinity, whereas the stellar radiations and winds from
 massive star(s) play a crucial role in the evolution of the nearby cloud and thus can induce next generation star formation
(Elmegreen \& Lada 1977).  The minimum criteria to ascertain such processes is that the second generation star(s)
should have younger age in comparison to nearby massive star(s). The 
physical properties of the compact and \uchii~ regions in the gas ridge suggest 
that they are younger than adjacent \hii regions S255 and S257 (see Section 3).
The age of the sources inside the dense ridge is of the order 
of $\sim$ 1.2 Myr or less and considering 1.6 Myr as a 
typical value for the age of evolved \hii regions, it appears that the 
sources in the gas ridge are younger than the adjacent massive stars 
as well as lower mass stars of the complex.
These signatures indicate that the sources found in the gas bar might have formed due to the interaction of the evolved \hii regions. 

\subsection{Triggered Star Formation}

A number of mechanisms by which massive stars can affect
the subsequent star formation in the region have been proposed.
One such process is compression of pre-existing molecular clumps due to
impinging of ionization/shock front, leading to density enhancement, which when 
it exceeds the
local critical mass, collapse to form new stars 
(Bertoldi 1989; Lefloch \& Lazareff 1994). The observational
signature of such a process would be the limb-brighten rim, having a dense head 
and a tail, extending
away from the ionizing sources (e.g., Miao et al. 2009). The morphology of the radio emissions and/or
the H$\alpha$ emission around the sources S255-2a and S255-2b show symmetrical distribution, not like the
cometary morphology with a bright rim as found in case of bright-rimmed clouds (e.g.  Urquhart et al. 2006; Ogura \& Sugitani 1998).
The mm contours also do not show any asymmetric distribution of molecular matter, 
as expected in case of the cometary morphology (e.g. Morgan et al. 2004).
The overall distribution of molecular material and
the distribution of radio continuum do not give strong 
evidence in favour of star formation
due to compression of pre-existing molecular clumps. 
However, the compression of pre-existing filamentary 
cloud cannot be ignored. Another proposed mechanism
that can induce star formation around the \hii region is a ``collect \& collapse" process.
In this mechanism, the neutral matter piles up between the ionization  and the
shock fronts of the expanding \hii region.  This compressed  material may be dynamically unstable
and fragments into dense clumps which can then form new stars (Deharveng et al. 2005).
We observed recent star formation activity at the interface of the S255 and 
S257. If the \hii regions are evolved in the same ambient medium, and S255 
is being excited by a relatively early type source, then we expect to 
have a larger impact on the star formation process at the interface. 
Therefore, we 
compare the 
observed properties with that of the analytical model of star formation at the periphery of the expanding \hii 
region through ``collect and collapse'' process by Whitworth et al. (1994).
The model predicts the time when the fragmentation starts ($t_{frag}$) and 
the radius at the time of
fragmentation ($R_{frag}$), which weakly depend upon isothermal sound speed 
in the compressed layer ($a_s$), and
varies from 0.2 - 0.6~km s$^{-1}$,  the number of Lyman continuum 
photons ($Q_{Ly}$) emitted by the massive
star and the density of the ambient medium ($n_0$) in which the 
expansion of \hii region occurs and are given by:
\begin{eqnarray}
\label{eq_rfrag}
R_{frag}&{\simeq}&5.8~a_{.2}^{4/11}~Q_{49}^{1/11}~n_3^{-6/11}~pc\\
\label{eq_tfrag}
t_{frag}&{\simeq}&1.56~a_{.2}^{7/11}~Q_{49}^{-1/11}~n_3^{-5/11}~Myr,
\end{eqnarray}
where $a_{.2}$=[$a_s$/0.2~km s$^{-1}$], $Q_{49}$=[$Q_{Ly}$/10$^{49}$
s$^{-1}$] and $n_3$=[$n_0$/10$^{3}$ cm$^{-2}$] are dimensionless
variables. Adopting 0.2 \kms for $a_s$, 2.4 $\times$ 10$^{48}$ photons s$^{-1}$ 
for $Q_{Ly}$ for the ionizing star 
(O9.5V) of S255 (Vacca et al. 1996) and 3 $\times$ 10$^{3}$ for 
$n_0$ (Chavarr\'{\i}a et al. 2008 and references therein), 
we obtained $t_{frag}$ \app 1.1 Myr and $R_{frag}$ \app 2.8 pc.
This indicates that the fragmentation will start at a radius of 2.8 pc, whereas
the young cluster and associated stellar sources are situated at a 
radius of $\sim$ 1.7 pc from the ionizing
star of S255. If one considers the present age of the YSOs ($\sim$ 1.2 Myr) 
plus the fragmentation time ($\sim$ 1.1 Myr),
one would expect an age difference of $>$ 2 Myr between the ionizing source of S255 (age $\sim$ 1.6 Myr) and the
YSOs (age $\sim$ 1.2 Myr) in the core. At the same time, to reach from the fragmentation time of the shell to
the present age of the YSOs ($\sim$ 1.2 Myr), one would expect a shell 
radius much beyond 2.8 pc.
Though the above estimations are based on the assumptions on the sound speed 
in the compressed layer. Increasing the value of sound speed increases 
the fragmentation radius and time further. It appears that the \hii region 
is not old enough to produce the S255-IR cluster by ``collect and collapse'' process.
Morphologically also, it seems that  ``collect and collapse'' process is not viable here,
as the cluster S255-IR is not associated with the  ring of PAH emissions that surround the \hii regions S255 or S257,
which generally represent the compressed neutral matter piled up between the ionization 
and shock fronts. Rather, the cluster is situated at the tip of the ring of PAH emissions. For example, in
the case of a sample of \hii regions by Deharveng et al. (2005), where the star formation at their peripheries is expected
due to ``collect and collapse'' process, one can see young 
sources/condensations, which are associated with the
ring of PAH emissions that delineate the \hii regions. Also in an ideal case,
one would expect regularly spaced  clumps along the periphery of \hii regions and/or young protostars of roughly  
similar ages, distributed preferentially along the swept-up material, 
which is not the case here,
whereas we are seeing sources at different evolutionary stages across the gas ridge and dense clumps. The radio emission
from S255 and S257 at 610 MHz is rather spherical and does not show
morphology similar to  the Champagne flow or blister type \hii regions, where one would expect the evolution of the
\hii region in a non-uniform medium. As a result, clump formation due to ``collect and collapse'' process may  
not be uniformly distributed. This discussion indicates that the ``collect and collapse'' process is also not at work here.

In the gas ridge which scenario should work better is difficult to conclude, 
but the existence of molecular clumps is not unique to the ``collect \& collapse'' 
model. Moreover, in this process the clumps should be distributed
in a ring like structure in the immediate \hii regions, not like the
vertical bar structure which we are witnessing here. 
The morphological similarity of the matter sandwiched between S255-S257 resembles 
on a large scale the process of star formation that has been observed
in the dense bar like structure in the Large Magellanic Cloud at the
interfaces of two expanding supergiant shells LMC4  and LMC5 (Cohen et al. 2003),
where it is believed that the collisions between the dense swept up neutral 
materials lead to star formation (Yamaguchi et al. 2001a,b). This similar kind 
of star formation activity might have happened here.
The above discussion reveals that it is difficult to constrain a particular scenario that is at work with the present analysis,
though we favour the last scenario. However, to strengthen the notion that the RDI and 
``collect \& collapse'' scenarios are not at work, it would be very useful 
to get radial velocity information and the age of the stellar members 
using spectroscopic
observations to confirm their association and relative evolutionary status, 
in order to obtain a better picture of triggered star formation.
  
\section {Conclusions}

1. Deep $JHK_{\rm s}$ observations have been carried out in a region 
of 4$^{\prime}$.9 $\times$ 4$^{\prime}$.9 centered on the S255-IR cluster 
associated with the gas ridge between two evolved \hii regions 
S255 and S257. Using the NIR CC and CM diagrams as main diagnostics of 
the PMS nature of the objects, we identified 109 likely YSO candidates 
to a mass limit of 0.1 \msun, which include 69 new YSO candidates in
the region observed by us. Our observations increased the number of 
previously identified YSOs in this region by 32\%.

2. Using $J-H/J$ CM diagram and assuming an age of 1 Myr, we 
found that the masses of the YSOs are in the range of 0.1 - 4 \msun~
and they are embedded in the molecular cloud (mean \av~$\sim$ 10 mag).

3. Throughout the observed region, the distribution of YSOs with 
excess emission is not uniform and is more concentrated along the gas 
ridge found at the interface of the two \hii regions S255 and S257. The 
YSO surface number density is found to be maximum ($\sim$ 180 stars/pc$^2$) 
towards the cluster S255-IR, however one can also see sub-structures in 
the map, with a secondary peak ($\sim$ 87 stars/pc$^2$) towards the 
northern side of the S255-IR.

4. Out of three massive (M $\ge$ 100 \msun) dust continuum clumps along 
the gas ridge, we observed radio continuum emission in two clumps. The 
physical properties of the radio emission are similar to those of 
compact and \uchii~regions, in agreement with previous observations. 
The spectral types estimated from the Lyman continuum photons suggest 
that they are powered by massive B1-B2 type of stars. We suggest that 
massive star formation has started in two clumps and the massive young 
sources (IRS 1-3 and S255N) are at different evolutionary stages and 
are probably younger than the low mass YSOs of the region.

5. The slope of the KLF in the gas ridge is found to be 0.17 $\pm$ 0.03,
which is similar to that obtained for young clusters and indicates a
young age of $<$ 1 Myr for the stellar sources. However, there is an 
indication of a break in the power law at $K_{\rm s}$ = 15.25 mag. The KLF 
slope in the magnitude range of 12.5 - 15.25 can be represented by 
$\alpha$ = 0.25 $\pm$ 0.05 and the KLF slope is found to be flat in the 
magnitude range of 15.25 - 17.75.

6. We constucted $V-I/V$ CM diagram of the YSOs (distributed outside 
the gas ridge) identified on the basis of NIR and IRAC observations.
The positions of these YSOs in the CM diagram indicate approximate age between
0.1 - 4 Myr, suggesting a possibility of non-coeval star formation in the 
S255-S257 complex.

7. The evolutionary status of 31 YSO candidates located outside the 
gas ridge and 16 YSO candidates located within the gas ridge has been studied 
using the SED fitting models. The models predict that the mean age of the 
YSO candidates outside the gas ridge is $\sim$  2.5 Myr, accreting with 
disk accretion rate in the range of 10$^{-7}$ $-$ 10$^{-8}$ $M_\odot$ yr$^{-1}$, 
while the mean age of the YSO candidates inside the gas ridge is $\sim$ 1.2 Myr, 
accreting with disk accretion rate in the range of 
10$^{-6}$ $-$ 10$^{-7}$ $M_\odot$ yr$^{-1}$. This indicates that the YSOs 
inside the gas ridge are younger than those outside the gas ridge.

8. We find that the morphology of the molecular material and the distribution of YSOs
at the interface of two optically visible evolved \hii regions, resembles,
on a large scale, the star formation that has been observed at the interfaces
of two supergiant bubbles LMC4  and LMC5, where it is believed that the 
star formation has occurred due to the collision of the swept up material 
by the bubbles. Hence, we believe this may be the site of induced star 
formation as found in the case  of LMC4 and LMC5.

\acknowledgements
We thank the anonymous referee for a critical reading
of the paper and several useful comments and suggestions,
which greatly improved the scientific content of the paper. 
The authors thank the staff of HCT operated by Indian Institute
of Astrophysics (Bangalore), IGO operated by Inter-University Centre
for Astronomy \& Astrophysics (Pune), the University of Hawaii 2.2 m 
telescope for supporting the first run of SIRIUS and GMRT managed by 
National Center for Radio Astrophysics of the Tata Institute of 
Fundamental Research (Mumbai) for their assistance and support during 
observations. This research has also made use of the NASA/IPAC
Infrared Science Archive, which is operated by the Jet Propulsion 
Laboratory, Caltech, under contract with the NASA. 
This paper used data from the NRAO VLA Archive Survey (NVAS). 
The NVAS can be browsed through \mbox{http://www.aoc.nrao.edu/$\sim$vlbacald/}.
We thank Annie Robin for letting us use her model of stellar population 
synthesis. We thank Luis Chavarr\'{\i}a for providing us the table
of IR-excess YSOs.

%\newpage

\begin{table}
\caption{Log of observational data}
\begin{tabular}{llllll}
\hline
\hline
Date (UT)  & Object  &  Filter     & No of & Exp (sec) & Telescope \\
& & &\multicolumn{1}{c}{frames} & \multicolumn{1}{c}{per frame} \\
\hline
%%2004 Dec 15 & S255/257 &\hspace{2mm} U & \hspace{2mm}    2         &\hspace{4mm} 1200 & ST\\
2004 Dec 15 &S255/257 &\hspace{2mm} B & \hspace{2mm}    3         &\hspace{4mm} 600& ST\\
2004 Dec 15 &&\hspace{2mm} V & \hspace{2mm}    3         &\hspace{4mm} 600& ST\\
2004 Dec 15 &&\hspace{2mm} I$_{c}$ & \hspace{2mm}    3         &\hspace{4mm} 600& ST\\
2005 Oct 10 &&\hspace{2mm} H${\alpha}$ & \hspace{2mm}    1         &\hspace{4mm} 900& ST\\
%%2009 Oct 14 & S255/257    &\hspace{2mm}  U & \hspace{2mm}    2         &\hspace{4mm} 300& ST\\
2009 Oct 14  &S255/257 &\hspace{2mm} B & \hspace{2mm}    2         &\hspace{4mm} 200& ST\\
2009 Oct 14  &&\hspace{2mm} V & \hspace{2mm}    2         &\hspace{4mm} 150& ST\\
2009 Oct 14  &&\hspace{2mm} I$_{c}$ & \hspace{2mm}    2         &\hspace{4mm} 60& ST\\
%%2009 Oct 14 & SA98&\hspace{2mm} U & \hspace{2mm}    3         &\hspace{4mm} 600& ST\\
2009 Oct 14 &SA98 &\hspace{2mm} B & \hspace{2mm}    3         &\hspace{4mm} 600& ST\\
2009 Oct 14 &&\hspace{2mm} V & \hspace{2mm}    3         &\hspace{4mm} 600& ST\\
2009 Oct 14 &&\hspace{2mm} I$_{c}$ & \hspace{2mm}    3         &\hspace{4mm} 100& ST\\
2008 Jan 11 &S255&\hspace{2mm} IFOSC7 & \hspace{2mm}  1         &\hspace{4mm} 1200& IGO\\
2008 Jan 11 &S257&\hspace{2mm} IFOSC7 & \hspace{2mm}  1         &\hspace{4mm} 1200& IGO\\
2009 Dec 20 &S255-2a&\hspace{2mm} Gr 7 & \hspace{2mm}  1         &\hspace{4mm} 1800& HCT\\
2010 Jan 18 &S255-2b&\hspace{2mm} Gr 7 & \hspace{2mm}  1         &\hspace{4mm} 2400& HCT\\
2000 Oct 13 &S255-IR &\hspace{2mm} J & \hspace{2mm}    18         &\hspace{4mm} 20& UH\\
2000 Oct 13 &&\hspace{2mm} H & \hspace{2mm}    18       &\hspace{4mm} 20& UH\\
2000 Oct 13 &&\hspace{2mm} K & \hspace{2mm}    18        &\hspace{4mm} 20& UH\\
\hline
\end{tabular}
\end{table}
%%%%%%%%%%%%%%%%%%%%%%%%%
\begin{table}
\caption{Observed integrated flux and angular size for associated \hii regions at 14.94 GHz}
\begin{tabular}{cccccc}
\hline\hline
%Frequency &\multicolumn{2}{c}{} &\multicolumn{1}{c}{1280 MHz} &\multicolumn{1}{c}{1280 MHz} \\ % & 4860MHz* \\
%\hline
Source& \multicolumn{2}{c}{Peak Position} & Int. Flux  & Angular Size\\
  & \multicolumn{1}{c}{$\alpha$(J2000)}& \multicolumn{1}{c}{$\delta$(J2000)} & \multicolumn{1}{c}{ mJy} & \multicolumn{1}{c}{} \\
\hline
S255-2a  &06:12:55.02 &+17:59:29.06  &3.307& 5\arcs.9 $\times$ 3\arcs.9 \\
S255-2b  &06:12:54.91 &+17:59:21.21  &1.620&3\arcs.0 $\times$ 2\arcs.7 \\
S255-2c &06:12:54.08 &+17:59:24.20  &3.060&3\arcs.2 $\times$ 1\arcs.3  \\
S255-N &06:12:53.52 &+18:00:26.16  &9.259& 2\arcs.5 $\times$ 1\arcs.4 \\
\hline
\end{tabular}
 \end{table}
%%%%%%%%%%%%%%%%%%%%%%%%%%
\begin{table}
\caption{Derived physical parameters for the \hii regions from radio observations}
\begin{tabular}{cccccccc}
\hline\hline
Source &\multicolumn{1}{c}{$\tau_c$} &\multicolumn{1}{c}{rms $n_{e}$} &\multicolumn{1}{c}{EM} & \multicolumn{1}{c}{log N$_{Lyc}$} & ZAMS$\dag$ \\
  & \multicolumn{1}{c}{}
          & \multicolumn{1}{c}{(10$^{3}$ cm$^{-3}$)}
          & \multicolumn{1}{c}{(10$^{6}$ cm$^{-6}$ pc)} & \multicolumn{1}{c}{(ph. s$^{-1}$)} &
          \multicolumn{1}{c}{Spectral type}\\ \hline
\hline
S255-2a&0.0002&1.627&1.545&45.11 &B1\\
S255-2b &0.0002&2.522&2.185&44.77&B2\\

S255-2c &0.0009&5.637&7.891&45.07 &B1 \\

S255-N &0.003&11.210&28.555&45.55 &B1 \\

\hline
\end{tabular}
\flushleft\small{ $\dag$ From Panagia (1973)}
 \end{table}

%%%%%%%%%%%%%%%%%%%%%%%%%%
%\begin{table}
\begin{sidewaystable}
\centering
\scriptsize
\caption{Inferred physical parameters of the YSOs detected outside the gas ridge from SED fits}
\begin{tabular}{ccccccccccc}
\hline\hline
 ID & \multicolumn{1}{c}{RA} & \multicolumn{1}{c}{DEC} &\multicolumn{1}{c}{M$_{\ast}$} &\multicolumn{1}{c}{T$_{\ast}$ } &\multicolumn{1}{c}{t$_{\ast}$} & \multicolumn{1}{c}{  M$_{\rm disk}$ }
& \multicolumn{1}{c}{$\dot{M}_{\rm disk}$ } & \multicolumn{1}{c}{$\dot{M}_{\rm env}$ } & \multicolumn{1}{c}{A$_V$ } & $\chi^2_{\rm min}$ \\

  &\multicolumn{1}{c}{(J2000)} & \multicolumn{1}{c}{(J2000)} & \multicolumn{1}{c}{($M_\odot$)}
          & \multicolumn{1}{c}{(10$^{4}$ K)}
          & \multicolumn{1}{c}{(10$^{6}$ yr)} & \multicolumn{1}{c}{( $M_\odot$)} & \multicolumn{1}{c}{(10$^{-8}$ $M_\odot$/yr)} &  \multicolumn{1}{c}{(10$^{-6}$ $M_\odot$/yr )} &
          \multicolumn{1}{c}{mag}\\ \hline
\hline

1 & 93.309601 & 17.895531 &2.089	$\pm$	0.522 &7.642	$\pm$	2.087 &5.058	$\pm$	2.919 & 0.005	$\pm$	0.011 & 3.305	$\pm$	3.065 & 0.233	$\pm$	1.731 & 4.418	$\pm$	1.022 & 3.78\\
2 & 93.296382 & 17.914910 &2.023	$\pm$	0.526 &5.031	$\pm$	0.620 &2.076	$\pm$	1.985 & 0.011	$\pm$	0.017 & 2.761	$\pm$	2.523 & 0.235	$\pm$	0.744 & 3.017	$\pm$	0.449 & 3.68\\
3 & 93.269846 & 17.930154 &1.161	$\pm$	0.562 &4.349	$\pm$	0.519 &2.082	$\pm$	1.830 & 0.003	$\pm$	0.007 & 0.676	$\pm$	0.646 & 2.234	$\pm$	51.020 & 4.103	$\pm$	0.856 & 9.62\\
4 & 93.248081 & 18.017855 &2.544	$\pm$	0.745 &7.392	$\pm$	2.437 &3.671	$\pm$	2.335 & 0.011	$\pm$	0.027 & 8.455	$\pm$	7.998 & 2.523	$\pm$	11.920 & 6.701	$\pm$	1.297 & 6.29\\
5 & 93.248309 & 18.031191 &2.962	$\pm$	0.818 &9.344	$\pm$	3.698 &3.916	$\pm$	2.449 & 0.010	$\pm$	0.023 & 6.824	$\pm$	6.509 & 3.625	$\pm$	30.850 & 6.100	$\pm$	1.084 & 4.75\\
6 & 93.242913 & 18.043001 &2.396	$\pm$	0.478 &5.454	$\pm$	1.749 &1.281	$\pm$	1.040 & 0.002	$\pm$	0.007 & 0.468	$\pm$	0.461 & 0.488	$\pm$	1.667 & 3.110	$\pm$	0.736 & 7.81\\
7 & 93.242141 & 18.073953 &0.931	$\pm$	0.471 &4.170	$\pm$	0.329 &1.446	$\pm$	0.757 & 0.001	$\pm$	0.003 & 0.132	$\pm$	0.161 & 2.134	$\pm$	45.570 & 3.054	$\pm$	0.549 & 3.40\\
8 & 93.233942 & 18.051073 &2.968	$\pm$	0.664 &9.488	$\pm$	2.604 &4.572	$\pm$	2.942 & 0.010	$\pm$	0.032 & 7.818	$\pm$	8.602 & 7.459	$\pm$	38.170 & 6.374	$\pm$	1.138 & 4.82\\
9 & 93.223580 & 18.021441 &3.011	$\pm$	0.615 &11.250	$\pm$	2.263 &5.765	$\pm$	2.013 & 0.002	$\pm$	0.007 & 2.169	$\pm$	2.493 & 0.005	$\pm$	0.077 & 7.262	$\pm$	0.624 & 1.57\\
10 &93.227723 & 18.041165 & 1.871	$\pm$	0.583 &5.759	$\pm$	1.848 &3.599	$\pm$	3.067 & 0.007	$\pm$	0.013 & 2.584	$\pm$	2.511 & 0.537	$\pm$	4.343 & 4.763	$\pm$	0.940 & 5.36\\
11 &93.235539 & 18.076643 & 1.504	$\pm$	0.424 &4.723	$\pm$	0.479 &3.042	$\pm$	2.250 & 0.004	$\pm$	0.008 & 0.422	$\pm$	0.398 & 1.186	$\pm$	39.370 & 3.734	$\pm$	0.542 & 5.84\\
12 &93.192612 & 17.948321 & 1.748	$\pm$	0.777 &4.567	$\pm$	0.491 &0.707	$\pm$	0.704 & 0.006	$\pm$	0.014 & 1.571	$\pm$	1.485 & 2.725	$\pm$	14.350 & 3.390	$\pm$	0.624 & 4.79\\
13 &93.224064 & 18.086501 & 1.571	$\pm$	0.493 &4.970	$\pm$	1.479 &2.363	$\pm$	1.938 & 0.001	$\pm$	0.003 & 0.036	$\pm$	0.047 & 0.583	$\pm$	24.560 & 3.195	$\pm$	0.872 & 2.42\\
14 &93.190166 & 17.949554 & 1.216	$\pm$	0.700 &4.257	$\pm$	0.442 &0.622	$\pm$	0.589 & 0.003	$\pm$	0.008 & 0.761	$\pm$	0.740 & 3.769	$\pm$	13.420 & 4.099	$\pm$	0.974 & 3.85\\
15 &93.197295 & 18.036892 & 1.887	$\pm$	0.658 &4.721	$\pm$	0.478 &1.218	$\pm$	1.014 & 0.001	$\pm$	0.004 & 0.134	$\pm$	0.144 & 0.929	$\pm$	12.690 & 3.548	$\pm$	0.683 & 6.77\\
16 &93.193666 & 18.069832 & 2.044	$\pm$	0.547 &5.626	$\pm$	2.224 &2.218	$\pm$	2.251 & 0.001	$\pm$	0.004 & 0.446	$\pm$	0.550 & 0.158	$\pm$	0.946 & 3.115	$\pm$	0.915 & 5.33\\
17 &93.283778 & 17.981431 & 2.186	$\pm$	0.983 &5.235	$\pm$	1.119 &1.963	$\pm$	2.030 & 0.022	$\pm$	0.029 & 22.700	$\pm$	21.050 & 8.179	$\pm$	31.220 & 6.318	$\pm$	0.902 & 5.02\\
18 &93.266869 & 18.058787 & 1.789	$\pm$	0.545 &5.119	$\pm$	1.401 &2.315	$\pm$	2.069 & 0.001	$\pm$	0.007 & 0.160	$\pm$	0.163 & 0.488	$\pm$	11.720 & 2.702	$\pm$	0.857 & 7.70\\
19 &93.250234 & 18.065654 & 1.155	$\pm$	0.520 &4.357	$\pm$	0.492 &2.132	$\pm$	1.771 & 0.003	$\pm$	0.006 & 0.536	$\pm$	0.503 & 1.430	$\pm$	40.350 & 4.100	$\pm$	0.811 & 2.75\\
20 &93.236915 & 18.072354 & 1.157	$\pm$	0.439 &4.415	$\pm$	0.504 &2.911	$\pm$	2.463 & 0.003	$\pm$	0.005 & 0.623	$\pm$	0.589 & 0.130	$\pm$	1.057 & 3.701	$\pm$	0.706 & 7.67\\
21 &93.207396 & 18.078301 & 1.922	$\pm$	0.638 &5.837	$\pm$	1.864 &3.756	$\pm$	3.162 & 0.007	$\pm$	0.014 & 2.872	$\pm$	2.699 & 1.231	$\pm$	8.120 & 4.845	$\pm$	1.100 & 5.13\\
22 &93.158847 & 17.917292 & 2.593	$\pm$	1.016 &4.870	$\pm$	0.927 &0.466	$\pm$	0.530 & 0.026	$\pm$	0.035 & 22.490	$\pm$	20.780 & 51.130	$\pm$	125.800 & 4.292	$\pm$	1.122 & 3.17\\
23 &93.189721 & 18.055715 & 1.635	$\pm$	0.332 &4.946	$\pm$	0.765 &3.767	$\pm$	2.472 & 0.005	$\pm$	0.009 & 0.490	$\pm$	0.451 & 0.032	$\pm$	0.738 & 3.029	$\pm$	0.473 & 2.35\\
24 &93.152634 & 17.918401 & 2.797	$\pm$	0.744 &5.005	$\pm$	0.589 &0.775	$\pm$	0.664 & 0.028	$\pm$	0.041 & 10.700	$\pm$	10.080 & 8.291	$\pm$	35.320 & 5.701	$\pm$	1.271 & 5.53\\
25 &93.131845 & 17.931312 & 1.345	$\pm$	0.758 &4.421	$\pm$	0.579 &1.505	$\pm$	1.314 & 0.005	$\pm$	0.008 & 1.231	$\pm$	1.159 & 1.389	$\pm$	10.260 & 4.468	$\pm$	1.066 & 4.62\\
26 &93.315675 & 18.051981 & 1.686	$\pm$	0.525 &4.716	$\pm$	0.457 &1.823	$\pm$	1.501 & 0.001	$\pm$	0.005 & 0.144	$\pm$	0.153 & 0.756	$\pm$	14.140 & 3.591	$\pm$	0.654 & 6.53\\
27 &93.205126 & 18.064643 & 2.227	$\pm$	0.603 &7.665	$\pm$	2.379 &4.900	$\pm$	2.760 & 0.010	$\pm$	0.017 & 7.030	$\pm$	6.758 & 3.788	$\pm$	28.590 & 6.289	$\pm$	1.184 & 1.77\\
28 &93.126392 & 17.930826 & 2.996	$\pm$	0.886 &6.156	$\pm$	1.675 &2.112	$\pm$	2.108 & 0.015	$\pm$	0.023 & 6.593	$\pm$	6.322 & 5.581	$\pm$	22.520 & 4.009	$\pm$	1.268 & 13.66\\
29 &93.318921 & 17.935201 & 2.523	$\pm$	0.782 &4.918	$\pm$	0.487 &0.984	$\pm$	0.674 & 0.003	$\pm$	0.007 & 0.433	$\pm$	0.417 & 2.171	$\pm$	9.788 & 6.622	$\pm$	0.847 & 7.73\\
30 &93.174703 & 18.062276 & 1.487	$\pm$	0.439 &4.732	$\pm$	0.468 &3.549	$\pm$	2.524 & 0.005	$\pm$	0.009 & 0.770	$\pm$	0.722 & 0.159	$\pm$	2.644 & 3.599	$\pm$	0.618 & 5.38\\
31 &93.132735 & 17.897481 & 2.939	$\pm$	0.520 &8.590	$\pm$	2.235 &2.630	$\pm$	1.010 & 0.008	$\pm$	0.025 & 11.320	$\pm$	10.910 & 1.551	$\pm$	13.240 & 4.397	$\pm$	0.623 & 2.92\\
\hline
\end{tabular}
% \end{table}
\end{sidewaystable}
%---------------------------------------------------
%\begin{table*}
\begin{sidewaystable}
\centering
\scriptsize
\caption{Inferred physical parameters of the YSOs detected inside the gas ridge from SED fits}
\begin{tabular}{ccccccccccc}
\hline\hline
 ID &\multicolumn{1}{c}{RA} & \multicolumn{1}{c}{DEC} &\multicolumn{1}{c}{M$_{\ast}$} &\multicolumn{1}{c}{T$_{\ast}$ } &\multicolumn{1}{c}{t$_{\ast}$} & \multicolumn{1}{c}{  M$_{\rm disk}$ }
& \multicolumn{1}{c}{$\dot{M}_{\rm disk}$ } & \multicolumn{1}{c}{$\dot{M}_{\rm env}$ } & \multicolumn{1}{c}{A$_V$ } & $\chi^2_{\rm min}$ \\

  & \multicolumn{1}{c}{(J2000)} & \multicolumn{1}{c}{(J2000)} & \multicolumn{1}{c}{($M_\odot$)}
          & \multicolumn{1}{c}{(10$^{4}$ K)}
          & \multicolumn{1}{c}{(10$^{6}$ yr)} & \multicolumn{1}{c}{( $M_\odot$)} & \multicolumn{1}{c}{(10$^{-8}$ $M_\odot$/yr)} &  \multicolumn{1}{c}{(10$^{-6}$ $M_\odot$/yr )} &
          \multicolumn{1}{c}{mag}\\ \hline
\hline

1 & 93.186031 & 17.951481 &0.720	$\pm$	0.775 &3.687	$\pm$	0.725 &1.392	$\pm$	2.219 & 0.006	$\pm$	0.013 & 5.980	$\pm$	7.953 & 17.490	$\pm$	88.960 & 5.561	$\pm$	1.291 & 1.69\\
2 & 93.254402 & 17.964297 &1.987	$\pm$	1.512 &4.789	$\pm$	1.812 &0.670	$\pm$	1.600 & 0.021	$\pm$	0.052 & 18.750	$\pm$	18.700 & 19.080	$\pm$	33.470 & 7.314	$\pm$	1.114 & 3.11\\
3 & 93.242582 & 17.980096 &3.085	$\pm$	0.819 &5.541	$\pm$	0.706 &1.188	$\pm$	0.620 & 0.007	$\pm$	0.016 & 1.568	$\pm$	1.430 & 27.540	$\pm$	92.210 & 9.688	$\pm$	1.637 & 2.71\\
4 & 93.237271 & 17.980611 &3.131	$\pm$	1.259 &5.468	$\pm$	2.361 &0.751	$\pm$	1.866 & 0.050	$\pm$	0.079 & 120.200	$\pm$	105.800 & 84.480	$\pm$	197.000 & 13.660	$\pm$	2.390 & 1.89\\
5 & 93.221573 & 17.984965 &3.024	$\pm$	1.152 &5.435	$\pm$	1.501 &0.974	$\pm$	1.195 & 0.026	$\pm$	0.036 & 78.710	$\pm$	72.440 & 20.010	$\pm$	55.650 & 10.690	$\pm$	1.938 & 4.54\\
6 & 93.249381 & 17.990787 &1.236	$\pm$	1.549 &3.770	$\pm$	0.716 &0.744	$\pm$	2.004 & 0.015	$\pm$	0.041 & 36.180	$\pm$	32.790 & 45.550	$\pm$	134.000 & 4.519	$\pm$	1.232 & 1.67\\
7 & 93.258103 & 17.998671 &0.950	$\pm$	0.879 &4.021	$\pm$	1.020 &1.040	$\pm$	1.873 & 0.007	$\pm$	0.010 & 5.771	$\pm$	5.241 & 8.943	$\pm$	25.210 & 6.680	$\pm$	1.622 & 3.30\\
8 & 93.222334 & 18.007723 &7.690	$\pm$	1.621 &8.597	$\pm$	4.189 &0.199	$\pm$	0.215 & 0.200	$\pm$	0.339 & 376.800	$\pm$	364.100 & 490.600	$\pm$	470.200 & 6.678	$\pm$	2.994 & 21.37\\
9 & 93.248134 & 18.017854 &2.537	$\pm$	0.602 &7.599	$\pm$	2.309 &4.045	$\pm$	2.705 & 0.015	$\pm$	0.026 & 6.645	$\pm$	6.236 & 2.594	$\pm$	17.380 & 6.067	$\pm$	1.379 & 3.81\\
10 &93.231215 & 18.020677 & 1.806	$\pm$	1.014 &5.177	$\pm$	1.952 &1.797	$\pm$	2.277 & 0.020	$\pm$	0.022 & 27.530	$\pm$	23.120 & 11.000	$\pm$	35.700 & 5.791	$\pm$	1.641 & 1.43\\
11 &93.223622 & 18.021473 & 2.620	$\pm$	0.990 &6.463	$\pm$	3.343 &1.500	$\pm$	2.165 & 0.010	$\pm$	0.022 & 3.900	$\pm$	3.650 & 9.623	$\pm$	31.410 & 7.729	$\pm$	1.217 & 3.85\\
12 &93.213193 & 18.023401 & 0.657	$\pm$	0.835 &3.588	$\pm$	1.041 &0.787	$\pm$	1.593 & 0.010	$\pm$	0.022 & 10.920	$\pm$	9.546 & 11.540	$\pm$	42.730 & 5.333	$\pm$	1.151 & 1.66\\
13 &93.222891 & 18.003234 & 3.636	$\pm$	0.959 &10.110	$\pm$	4.614 &2.773	$\pm$	2.665 & 0.015	$\pm$	0.034 & 11.770	$\pm$	11.330 & 16.330	$\pm$	60.620 & 22.140	$\pm$	2.431 & 2.08\\
14 &93.259425 & 18.004278 & 1.175	$\pm$	0.848 &4.310	$\pm$	1.007 &1.183	$\pm$	1.949 & 0.007	$\pm$	0.011 & 6.775	$\pm$	6.645 & 11.070	$\pm$	32.760 & 5.208	$\pm$	1.573 & 2.85\\
15 &93.213312 & 18.014792 & 1.193	$\pm$	1.341 &4.194	$\pm$	1.560 &0.321	$\pm$	1.316 & 0.014	$\pm$	0.032 & 35.600	$\pm$	42.540 & 15.450	$\pm$	24.570 & 5.882	$\pm$	1.219 & 1.05\\
16 &93.220701 & 17.996683 & 1.943	$\pm$	1.640 &4.210	$\pm$	0.337 &0.096	$\pm$	0.097 & 0.025	$\pm$	0.064 & 31.580	$\pm$	27.180 & 26.820	$\pm$	21.810 & 9.245	$\pm$	1.380 & 1.50\\
\hline
\end{tabular}
% \end{table*}
\end{sidewaystable}
%-------------------------

\begin{figure*}
 \vspace{-12cm}
\includegraphics[angle=0,width=18cm,height=26cm]{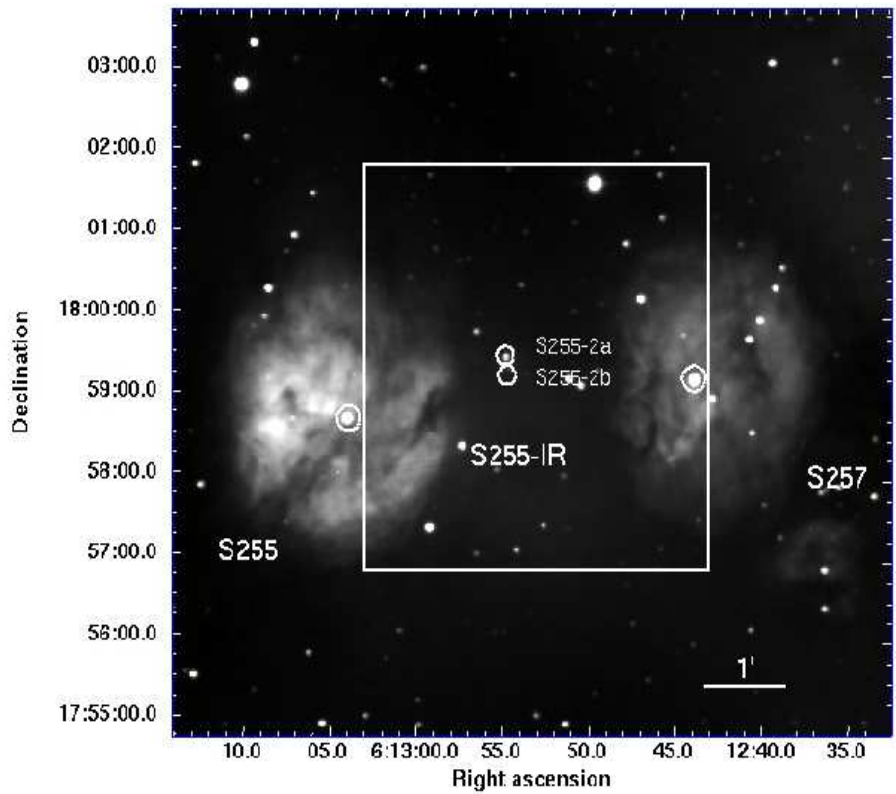}
\caption{H$\alpha$+continuum image of the ionized gas associated with
S255-S257 region obtained with ST (see Table 1) using the same set up described in Samal et al. (2007). The FOV is $\sim$ 10\arcm.0 $\times$ 9\arcm.5. 
The square box represents the observed region of the embedded cluster in
NIR bands. The individual \hii regions and the cluster S255-IR are also marked. The circles represent the ionizing
sources of the \hii regions, for which we have obtained optical spectroscopic information. 
North is up and east is to the left. RA and DEC coordinates are in J2000 epoch.}
\label{kdvel}
\end{figure*}

\begin{figure*}
\includegraphics[angle=0,width=8cm,height=12cm]{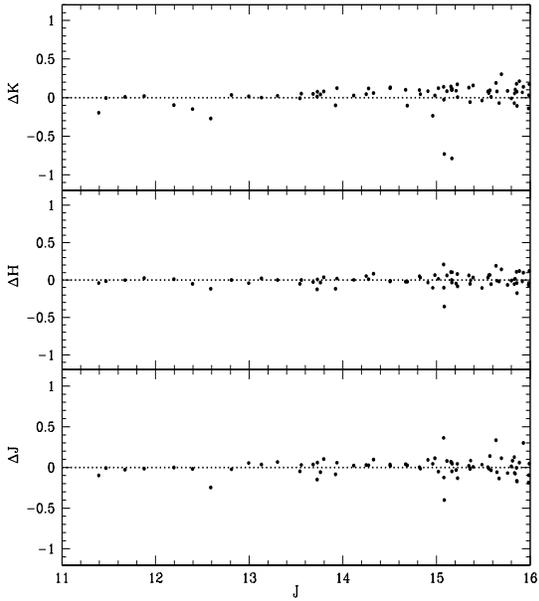}
% \vspace{1cm}
\caption{Comparison of UH/SIRIUS $JHK_{\rm s}$ photometry with the
2MASS data for the common sources. The
difference $\Delta$ (UH - 2MASS) in mag is plotted as a function
of $J$ mag.}
\label{kdvel}
\end{figure*}

\begin{figure*}
\includegraphics[angle=0,width=15cm,height=13cm]{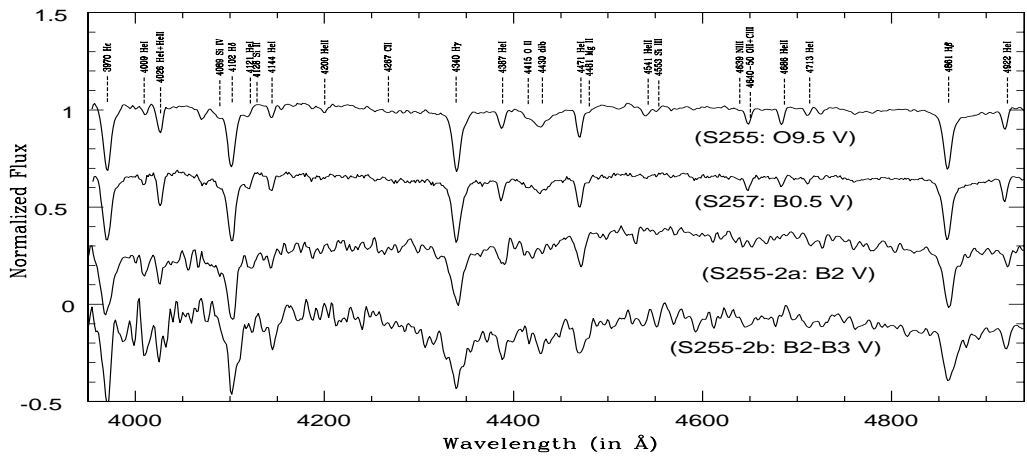}
% \vspace{1cm}
\caption{The blue part of the spectra of the ionizing sources of 
the \hii regions. Spectra are normalised 
and shifted with arbitrary units.}
\label{kdvel}
\end{figure*}

\begin{figure*}
\includegraphics[angle=0,width=9.0cm,height=11cm]{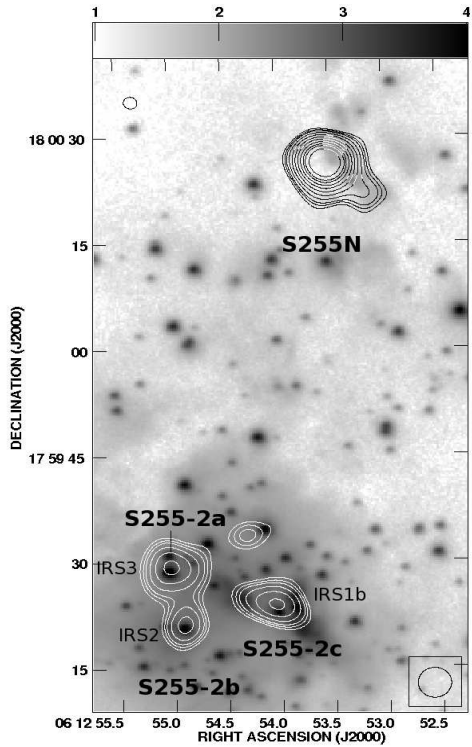}
% \vspace{1cm}
\caption{$K_{\rm s}$-band image in logarithmic scale overlaid with
the VLA radio continuum contours at 14.94 GHz. The contour levels are at
0.11 \into (3, 4, 5, 7, 10, 14, 19, 25, 32, 38, 50) mJy/beam,
where $\sim$ 0.11 mJy/beam is the rms noise in the map at a resolution
of $\sim$ $4\arcsec.4 \times 4\arcsec.2 $. 
The ionizing sources of the compact and \uchii~ regions
S255-2a, S255-2b and S255-2c, are labeled with IRS3, IRS2 and IRS1b 
respectively, for further discussion in the paper (see Section 4.1) }.
\label{kdvel}
\end{figure*}

\begin{figure*}
\includegraphics[angle=0,width=11.cm,height=15.5cm]{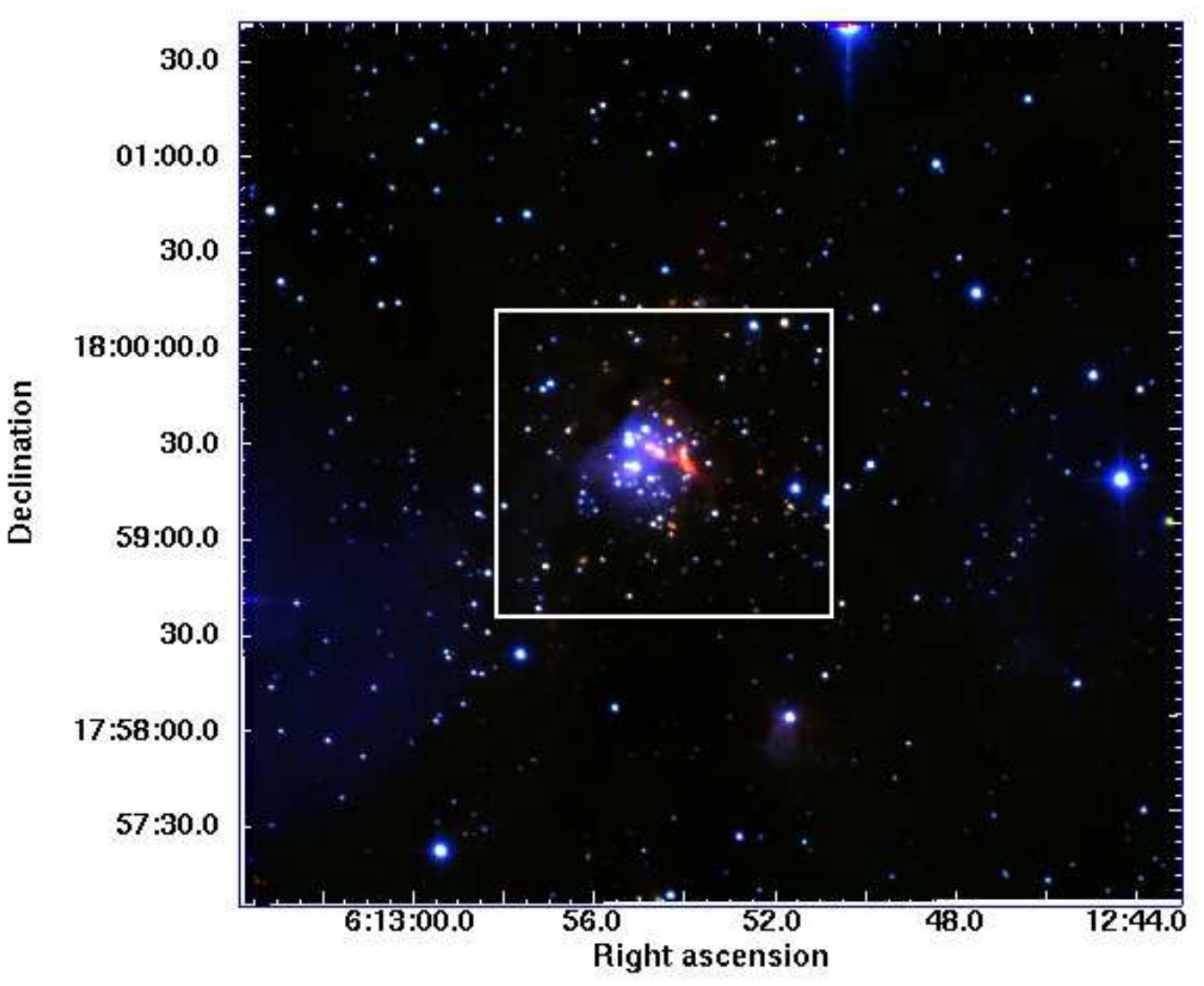}
\hspace*{-2cm}
\includegraphics[angle=0,width=10.0cm,height=14.4cm]{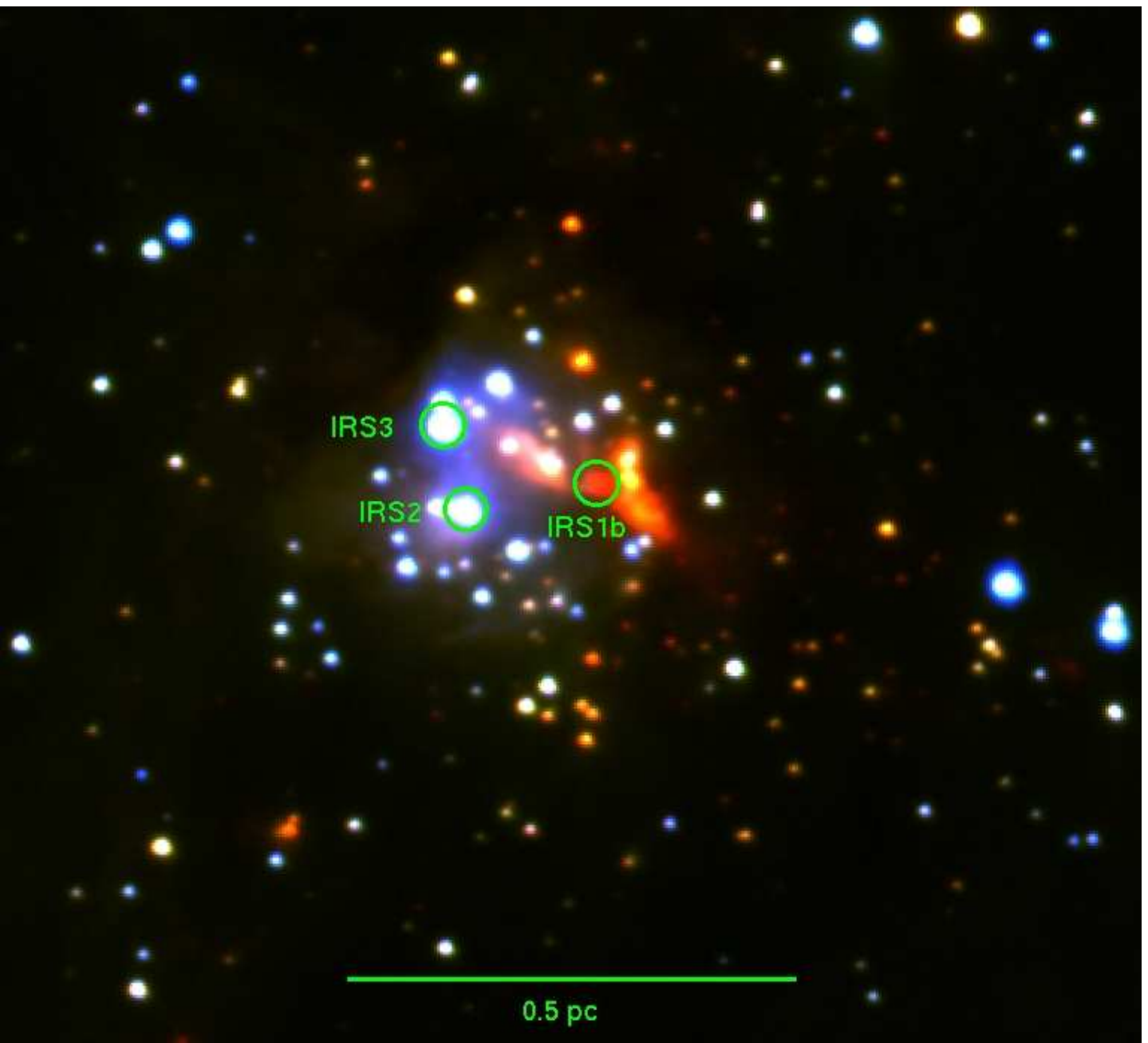}
% \vspace{1cm}
\caption{{\it Left panel}: \jhk color-composite image of the S255-IR SFR ($J$, blue; 
$H$, green; and $K_{\rm s}$, red). The FOV is $\sim$ 4\arcm.9 $\times$ 4\arcm.9. 
The box corresponds to
 the FOV ($\sim$ 1\arcm.5 $\times$ 1\arcm.5) 
of the color-composite image shown in the right panel. 
{\it Right panel}: The enlarged view of the color-composite image of the 
cluster S255-IR. The ionizing sources of the compact and \uchii~ regions 
S255-2a, S255-2b and S255-2c (marked with circles), 
are labeled with IRS3, IRS2 and IRS1b respectively.
North is up and east is to the left 
(see online electronic version for the color images).}
\label{kdvel}
\end{figure*}

\begin{figure*}
\includegraphics[angle=0,width=8.0cm,height=10.5cm]{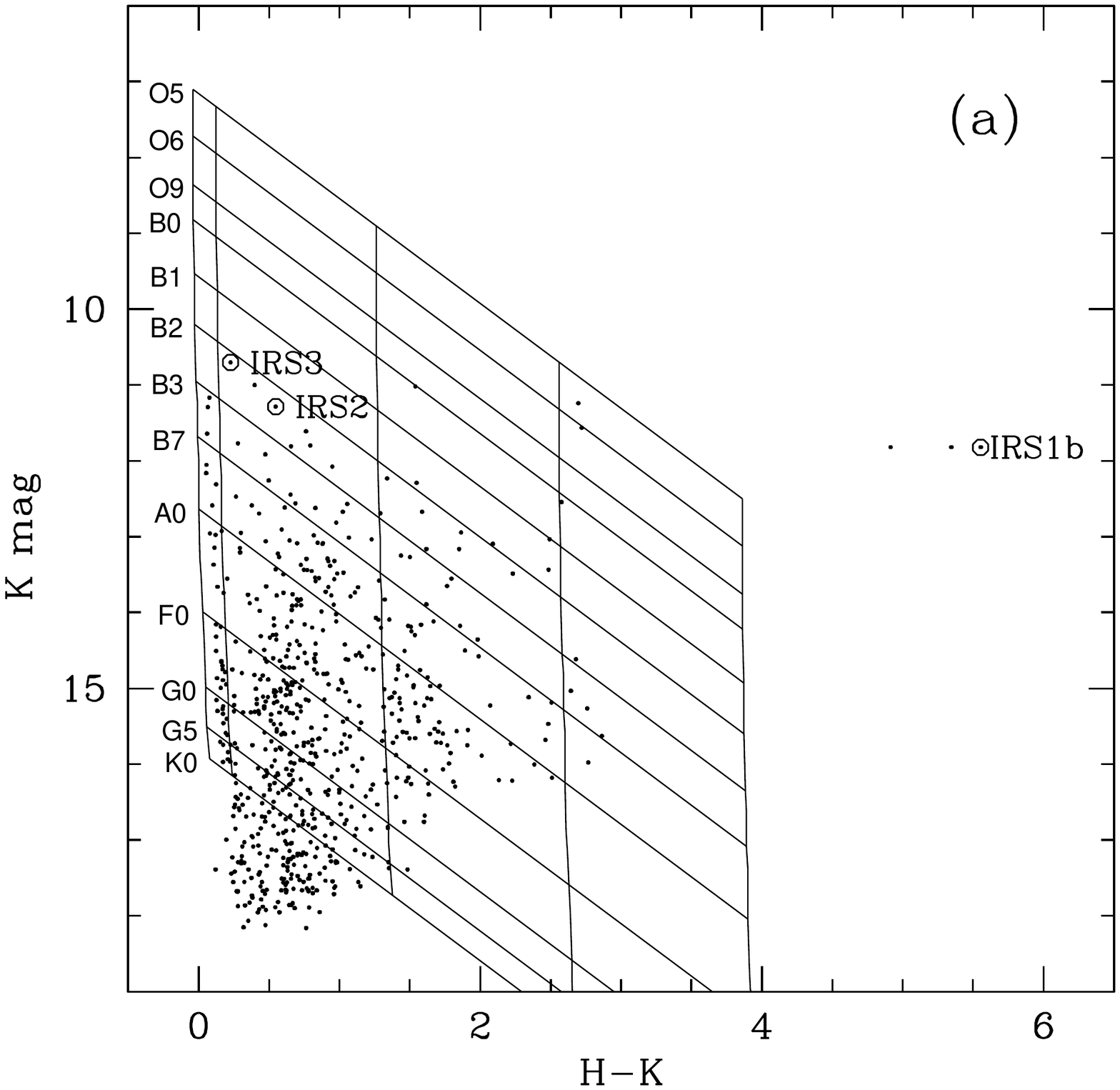}
\includegraphics[angle=0,width=8.0cm,height=10.5cm]{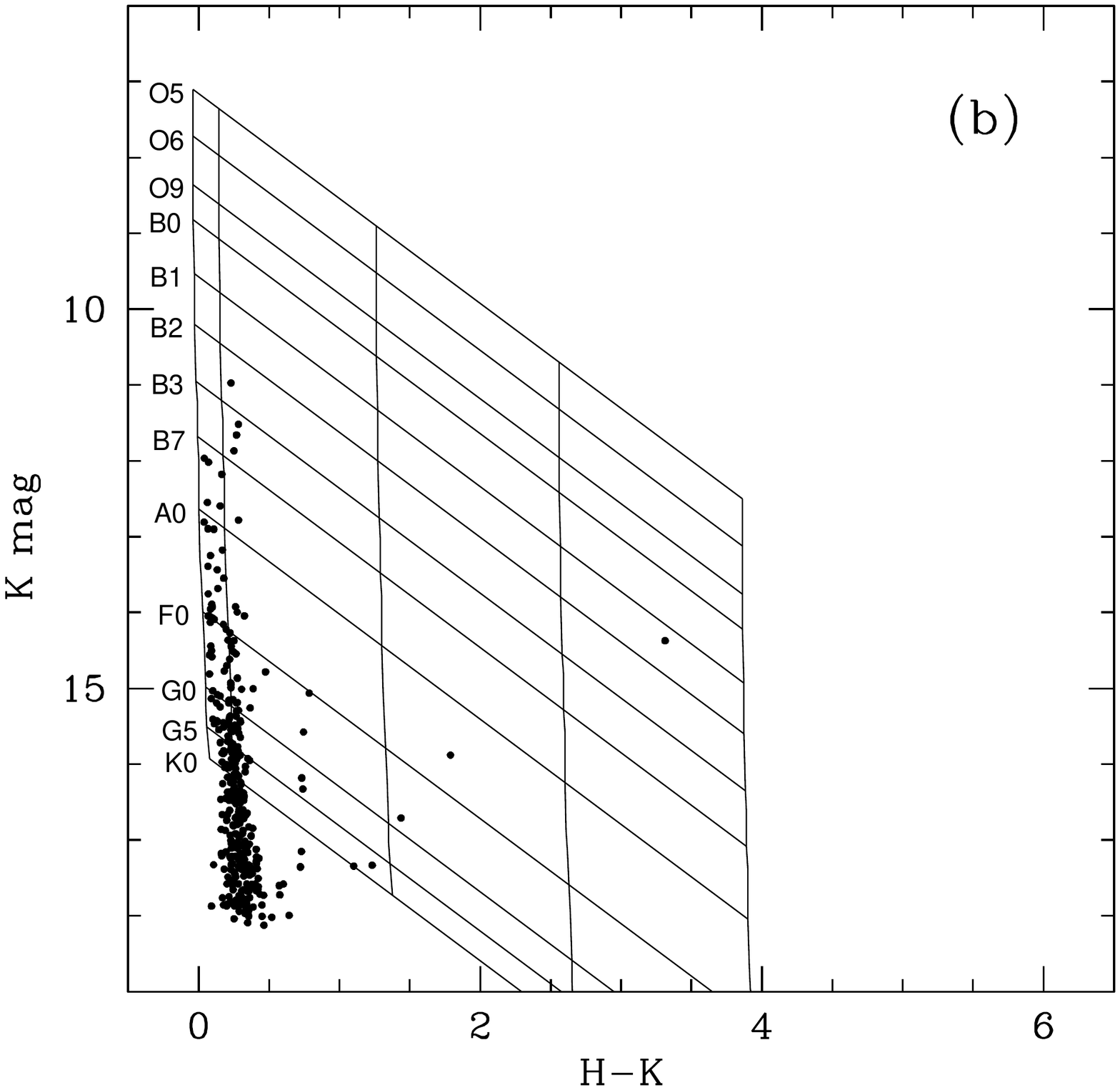}
\caption{(a) NIR CM diagram of the S255-IR SFR. The vertical 
solid lines from left to right indicate 
the MS track at 2.5 kpc reddened by \av = 0, 2.5, 20, 40, and 60 mag, respectively. 
Slanting horizontal lines represent the standard reddening vectors drawn from the MS locus
corresponding to different spectral types. The positions of the 
ionizing sources of S255-2a, S255-2b and S255-2c (shown as circles), are 
labeled with IRS3, IRS2 and IRS1b respectively. (b) NIR CM diagram for the control region.}
\label{kdvel}
\end{figure*}

\begin{figure*}
\includegraphics[angle=0,width=8.0cm,height=11.0cm]{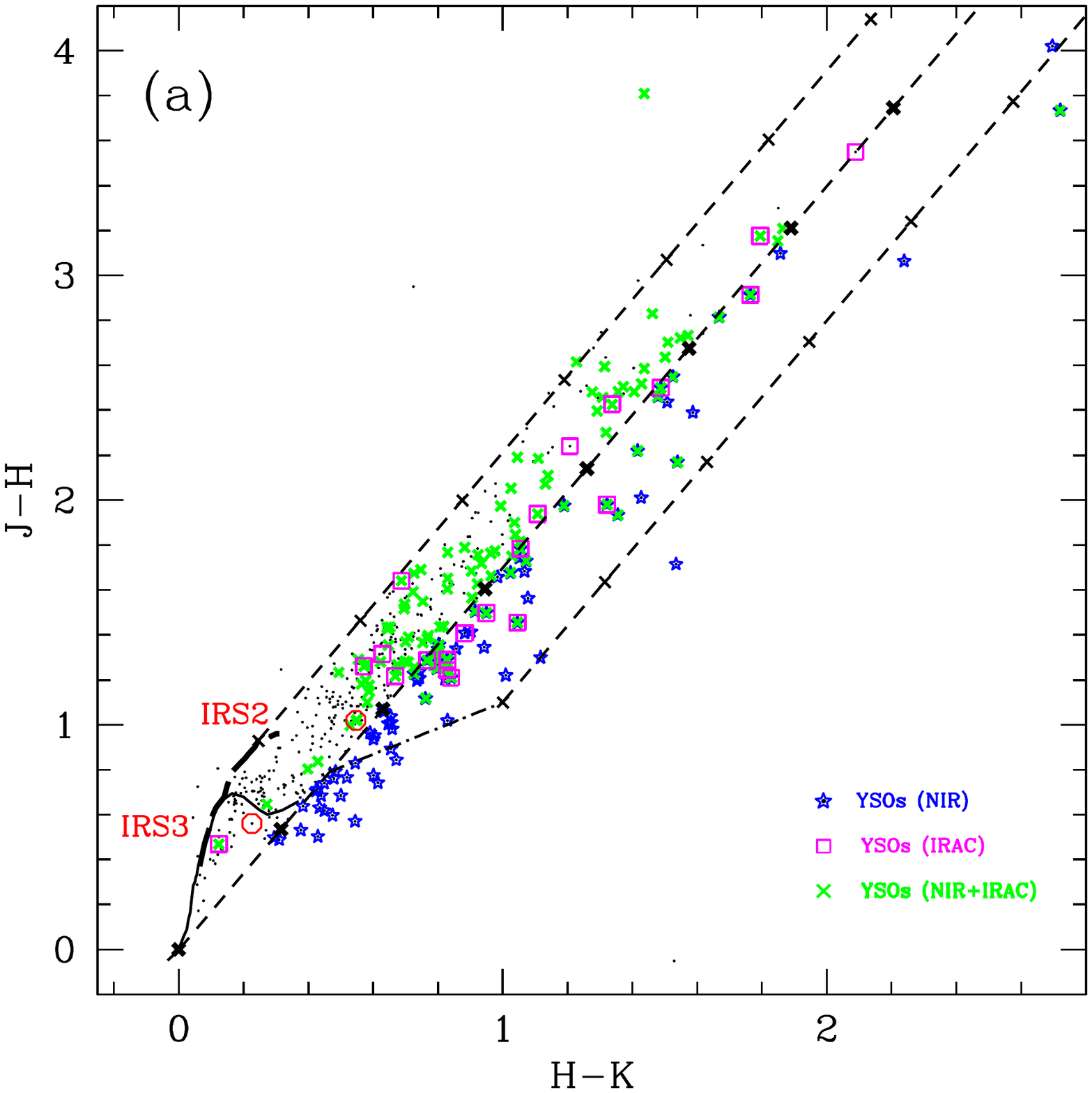}
\includegraphics[angle=0,width=8.0cm,height=11.0cm]{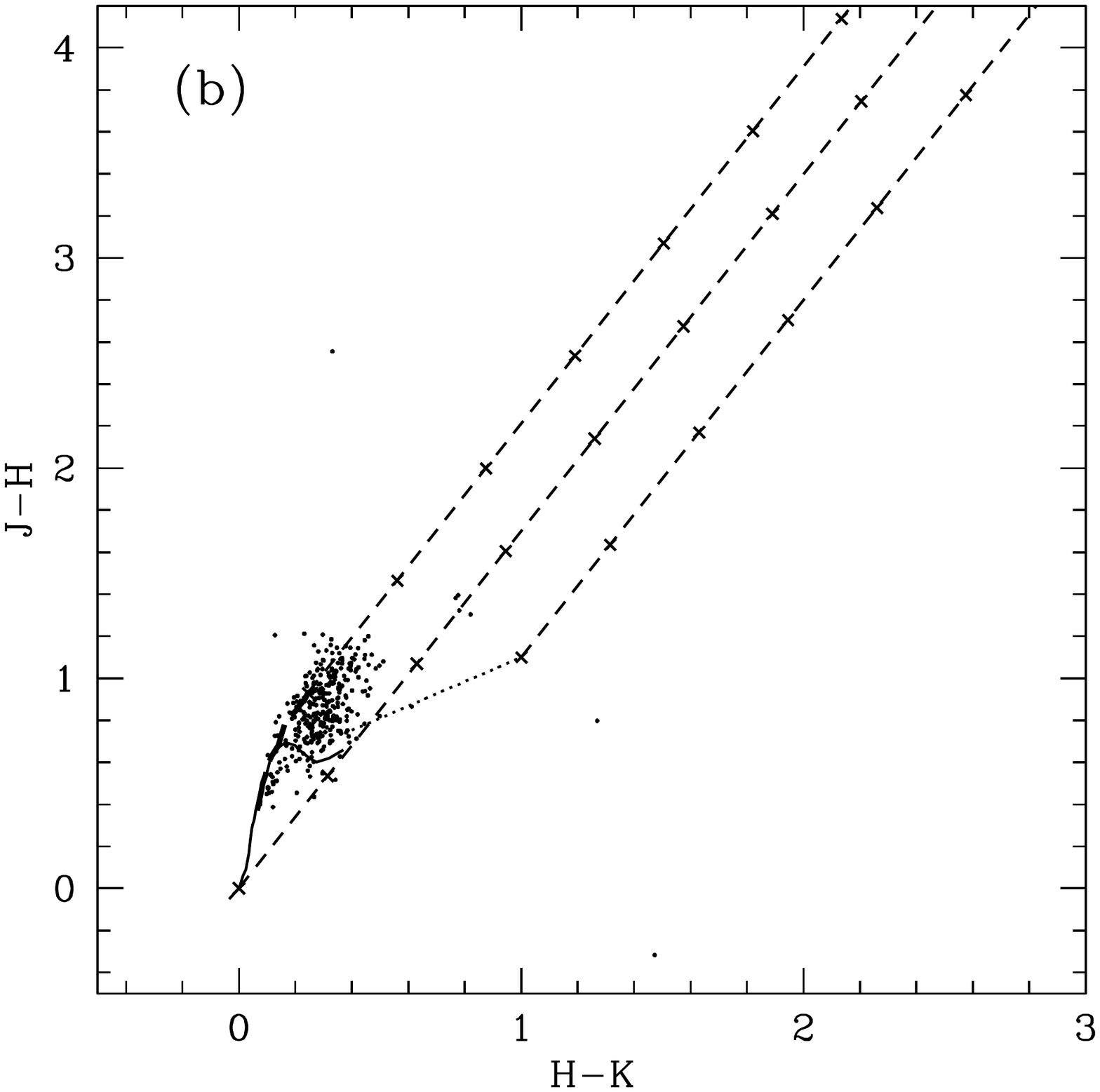}
% \vspace{1cm}
\caption{(a) NIR CC diagram of the S255-IR region for the sources
detected in the $JHK_{\rm s}$ bands (small dots). The continuous
line marks the locus of the  MS and thick dashed line is the locus of giant
stars. The three parallel dashed lines represent the reddening vectors, the
crosses represent a visual extinction of \av~= 5 mag. The locus of
CTTS is also shown with a dotted line.
The positions of the ionizing sources of S255-2a and S255-2b 
(shown as circles), are labeled with 
IRS3 and IRS2 respectively.
The squares are the IRAC identified Class II and Class I YSOs, whereas 
the crosses are the IR-excess YSOs identified using NIR plus IRAC bands
(for color plot see online electronic version). 
(b) NIR CC diagram for the control region.} 
\label{kdvel}
\end{figure*}

\begin{figure}
\includegraphics[angle=0,width=8cm,height=10.5cm]{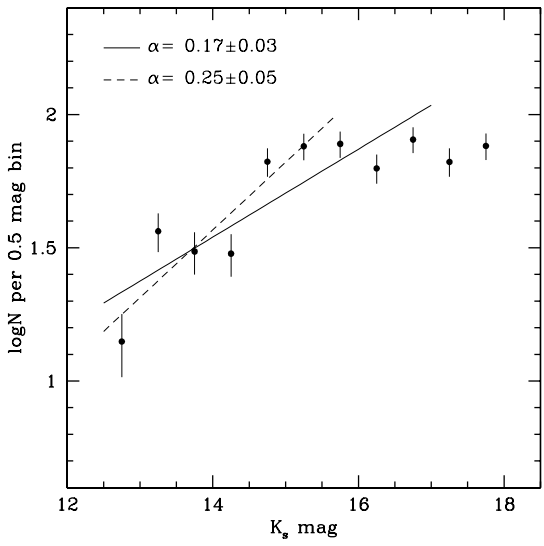}
% \vspace{1cm}
\caption{The completeness-corrected and field star-subtracted
KLF of the S255-IR region. The linear fits for various magnitude 
ranges are represented by the straight lines (see the text).
}
\label{kdvel}
\end{figure}

\begin{figure*}
\includegraphics[angle=0,width=13cm,height=17cm]{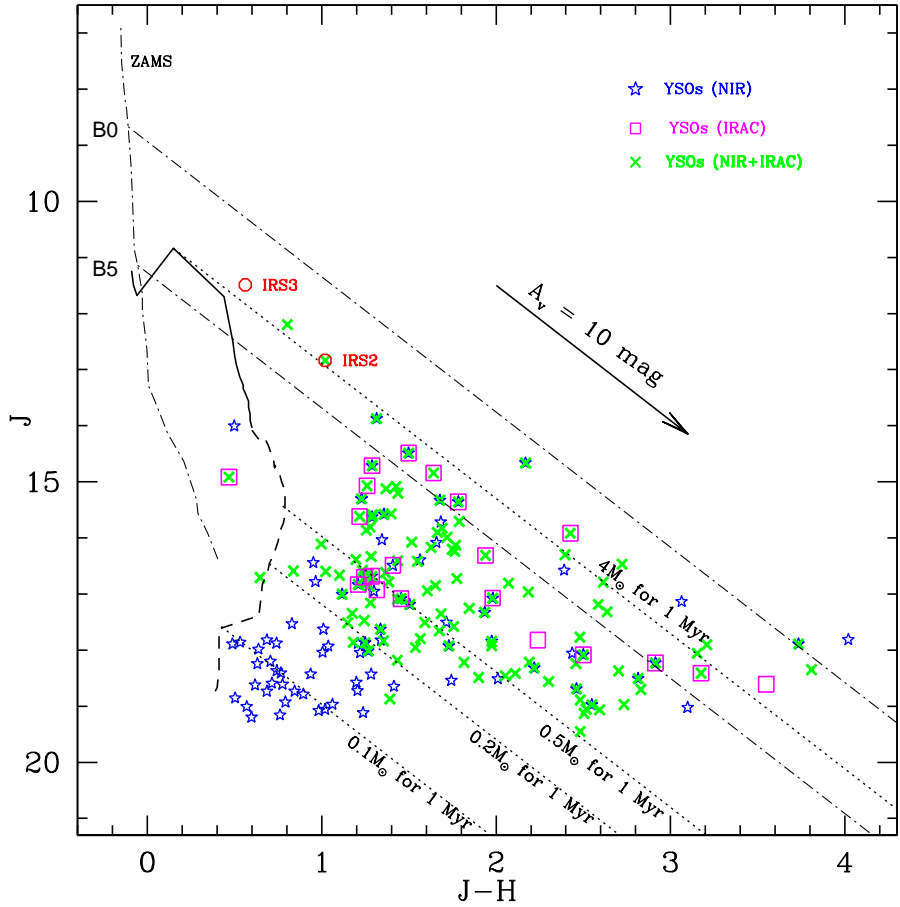}
\vspace{-1cm}
\caption{NIR CM diagram for YSOs in the S255 region. 
The asterisks represent the YSOs, identified on the basis of NIR CC 
diagram. The squares represent the sources that are identified 
as YSOs in the IRAC bands, whereas the crosses denote the IR-excess
YSOs identified using NIR plus IRAC bands.
The PMS isochrones of 1 Myr from Siess et al. (2000) and Baraffe et al. (1998) 
are drawn in solid and dashed curved lines, respectively,  
at a distance of  2.5 kpc and zero reddening. The reddening vectors 
corresponding to 0.1, 0.2, 0.5, and 4 $\msun$ are also shown in dotted slanting 
lines. The ZAMS, along with the reddening vectors (dashed-dotted slanting lines) 
from the tip of B0 and B5 stars are also shown. The ionizing sources of 
compact and \uchii~regions are marked in the same manner as in Figure 7 
(for color plot see online electronic version).
}
\label{kdvel}
\end{figure*}

\begin{figure*}
\vspace{-3cm}
\includegraphics[angle=0,width=13.0cm,height=16.5cm]{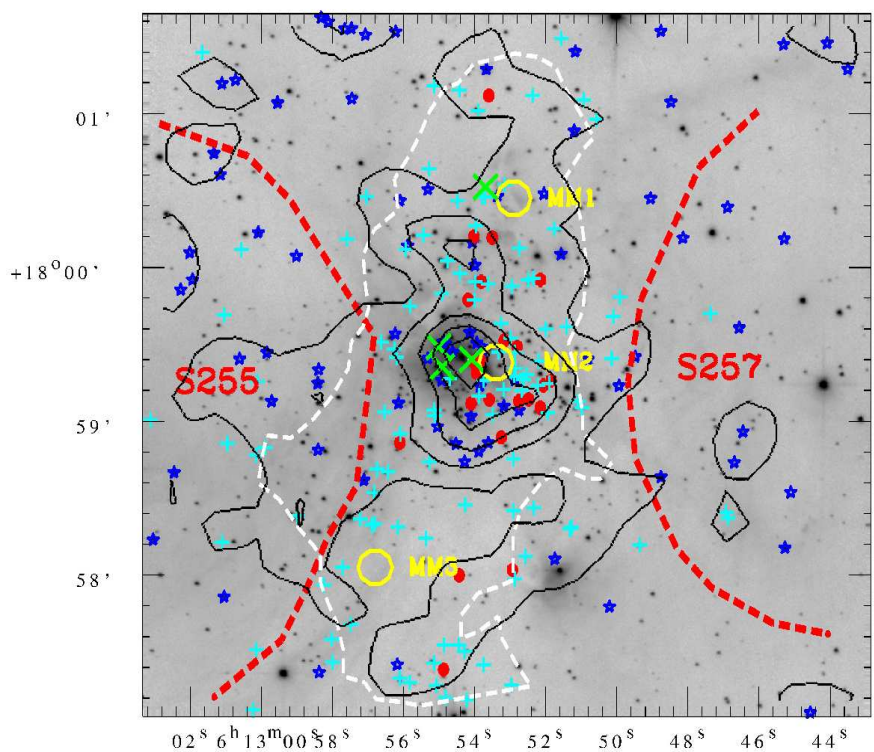}
\caption{The spatial distributions of YSOs identified using NIR CC 
diagram (asterisks), sources detected only in $H$ and $K_s$ bands
with $H-K$ $\geq$ 2 (filled circles) and the sources detected only in 
the $K_s$ band (plus symbols),
overlaid on \ksb image. The white dashed line shows the outer periphery of the dust 
continuum emission at 870 $\mu$m, which is at three times of the rms noise (501 mJy/beam) in the sub-milimeter map. The yellow circles are the positions of 1.2 mm continuum cores (MM1, MM2 and MM3; top to bottom) 
in the gas ridge. The red dashed 
lines are the demarcation of the outer boundaries of S255 and S257 \hii 
regions. The large cross signs represent the positions of the radio continuum sources. 
The YSO SSND contours are shown in black lines. The contour levels are drawn 
from 20 stars/pc$^2$ with an  increment of 30 stars/pc$^2$.
North is up and east is to the left (for color plot see online electronic version).
}
\label{kdvel}
\end{figure*}

\begin{figure}
\centering{
\includegraphics[angle=0,width=8cm,height=10cm]{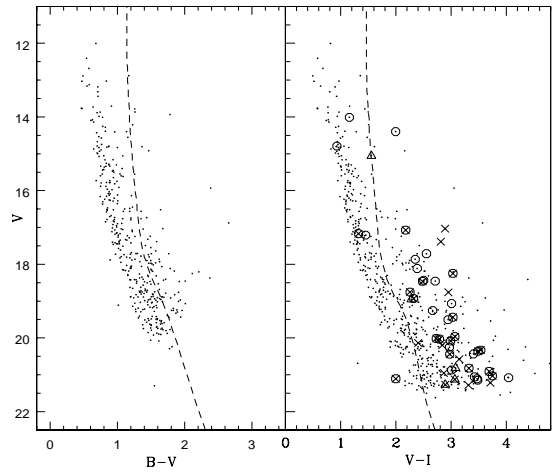}}
% \vspace{1cm}
\caption{The CM diagrams for stars lying in the S255-S257 cluster region.
The dashed curve represents ZAMS by Schmidt-Kaler (1982) corrected for 
$E(B-V)$ = 1.41 mag (\av~= 4.4 mag) and distance of 2.5 kpc. The 
Class II (circles), Class I (triangles) and IR-excess (crosses) YSOs 
identified by Chavarr\'{\i}a et al. (2008) are also marked.}
\label{kdvel}
\end{figure}

\begin{figure}
\centering{
\includegraphics[angle=0,width=12cm,height=15.5cm]{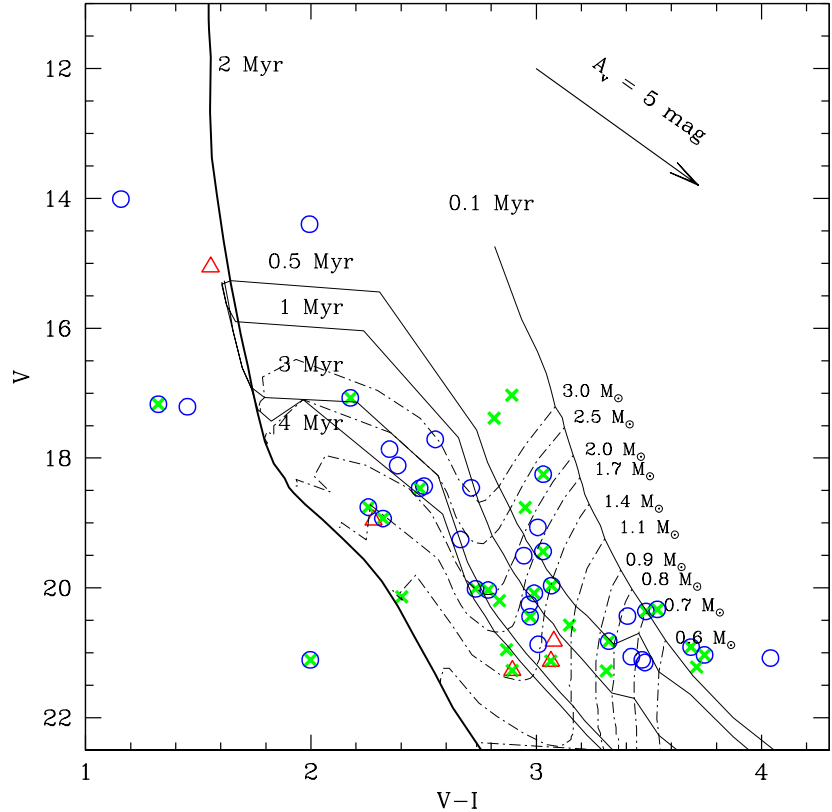}}
%\vspace{1cm}
\caption{Optical CM diagram for Class II (circles), 
Class I (triangles) and IR-excess (crosses) YSOs found by 
Chavarr\'{\i}a et al. (2008).
The isochrone for 2 Myr age for solar metallicity 
by Girardi et al. (2002) (thick curve) and PMS isochrones
of 0.1, 0.5, 1, 3 and 4 Myr (thin curves) along with evolutionary tracks 
of different mass (dashed-dotted lines) stars by Siess et al. (2000) 
are also shown. All the isochrones are corrected for the distance of 
2.5 kpc and reddening A$_V$ = 4.6 mag. 
The arrow indicates a reddening vector for A$_V$= 5 mag 
(for color plot see online electronic version).}
\label{kdvel}
\end{figure}

\clearpage
\begin{figure*}
\centering{
\includegraphics[width=5cm]{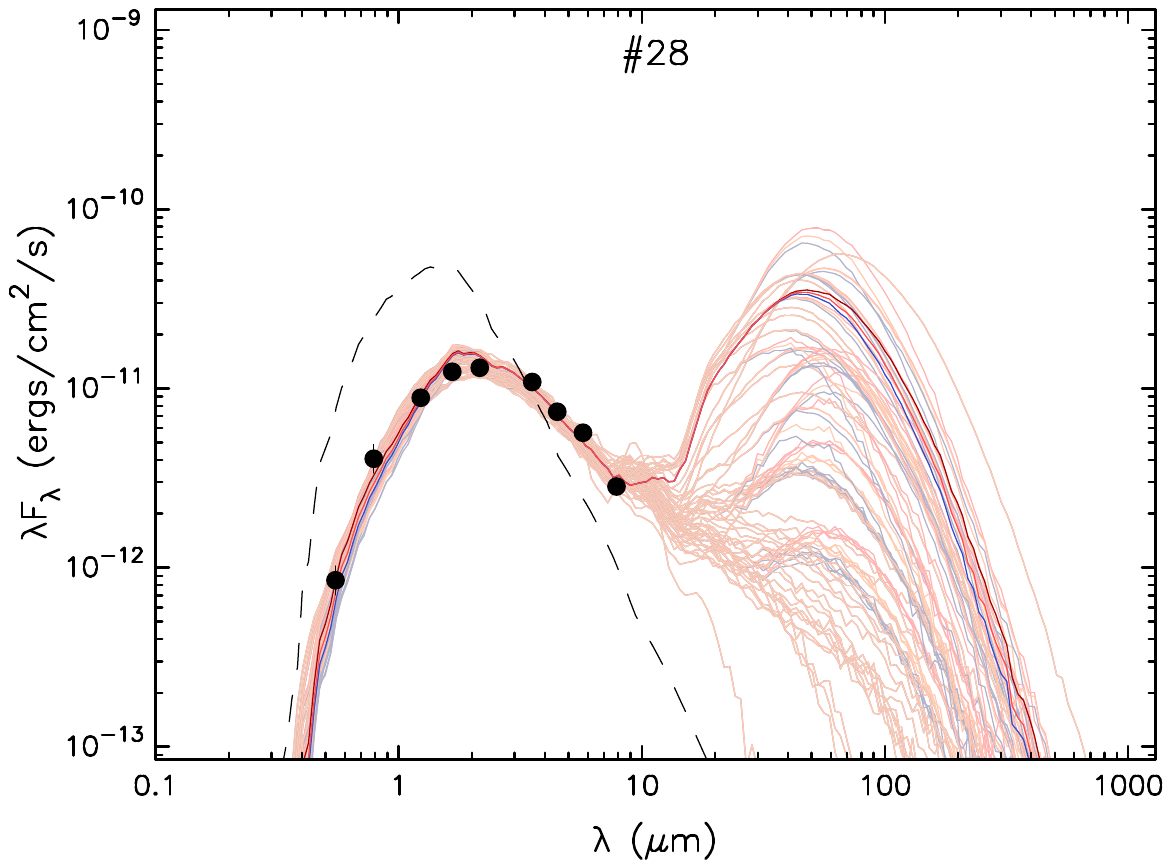}
\includegraphics[width=5cm]{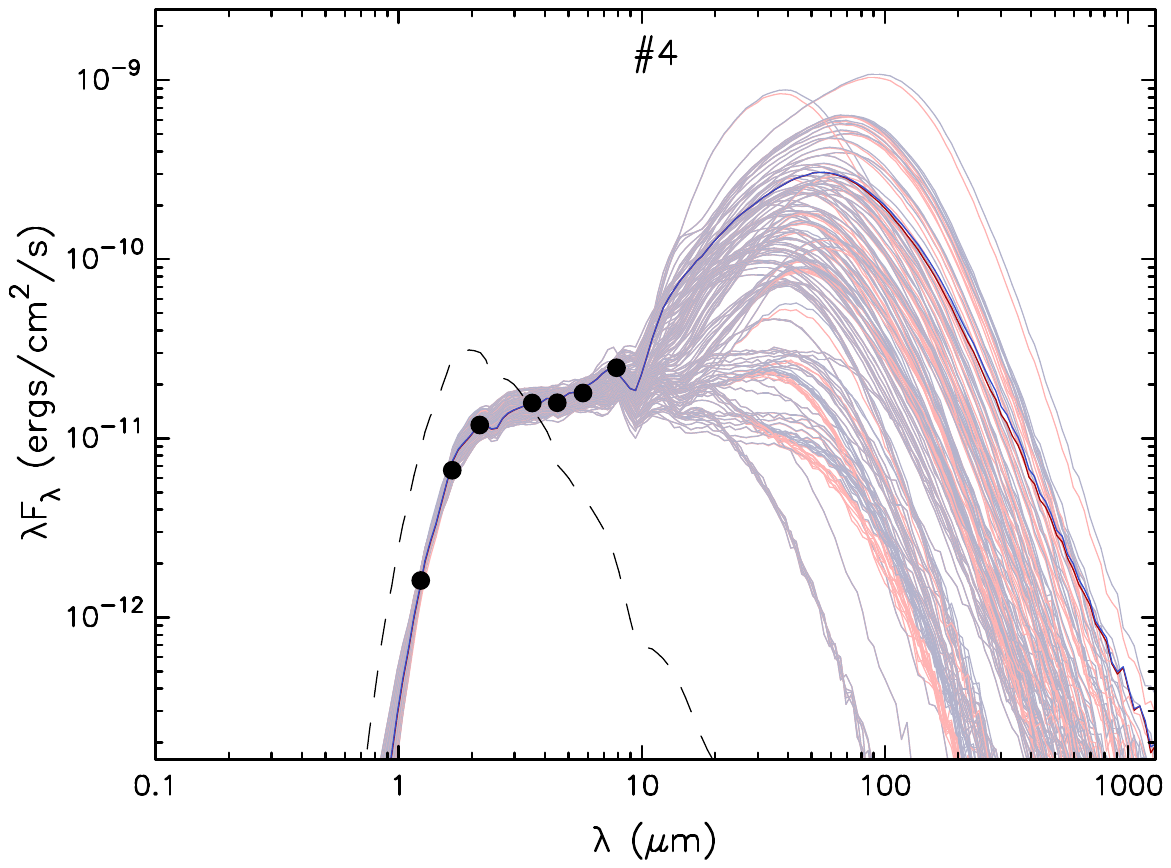}
\includegraphics[width=5cm]{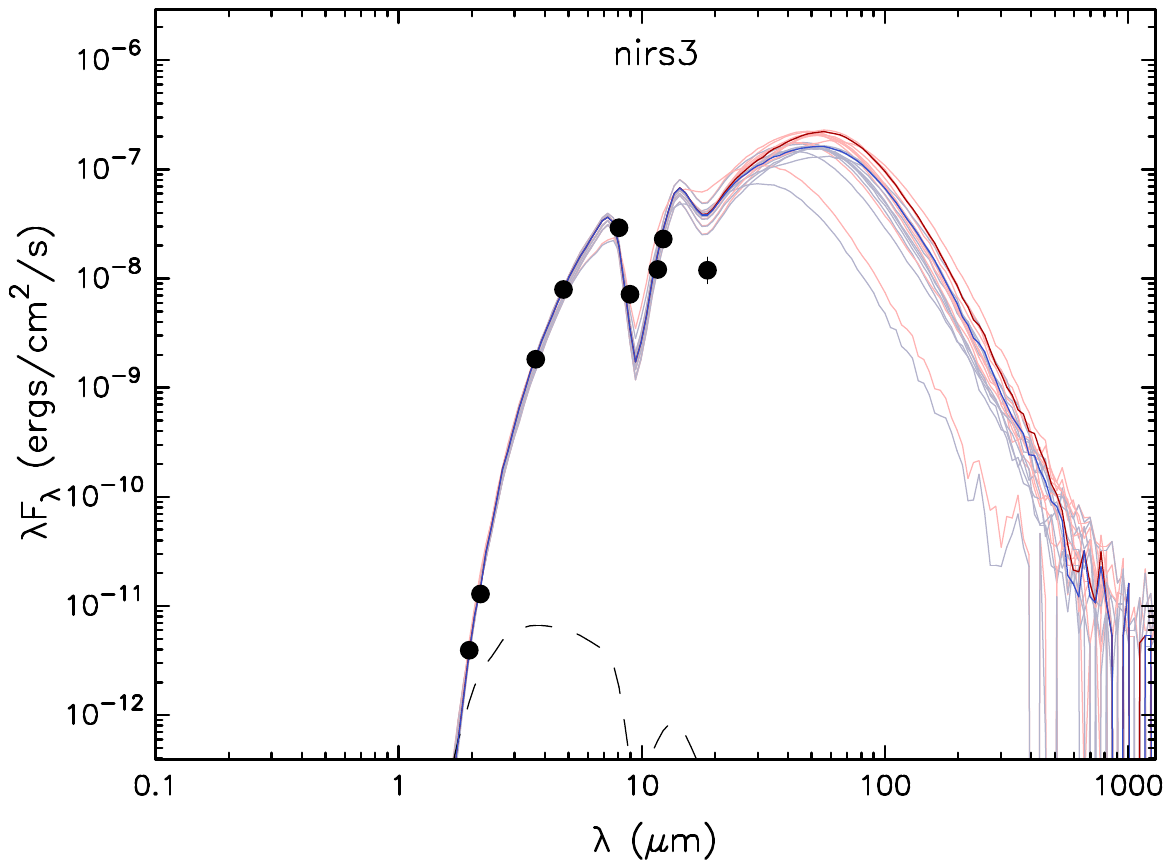}}
\caption{The SEDs of IRAC identified YSOs with optical and NIR counterparts 
outside the gas ridge (left), 
only NIR counterparts inside the gas ridge (middle) and the YSO 
associated with \uchii~ region S255-2c 
(right). The solid lines show the family of 
models that fit the data reasonably with 
${\chi}^2 - {\chi^2_{\rm min}} \leq 2N_{\rm data}$,
where ${\chi}^2_{\rm min}$ is the goodness-of-fit parameter for the
best-fit model (see the text).
The dashed line shows the SED of the stellar photosphere in the 
best-fitting model. The filled circles denote the
input flux values (for color plot see online electronic version).}
\label{kdvel}
\end{figure*}

%\clearpage
\begin{figure*}
\includegraphics[width=15cm]{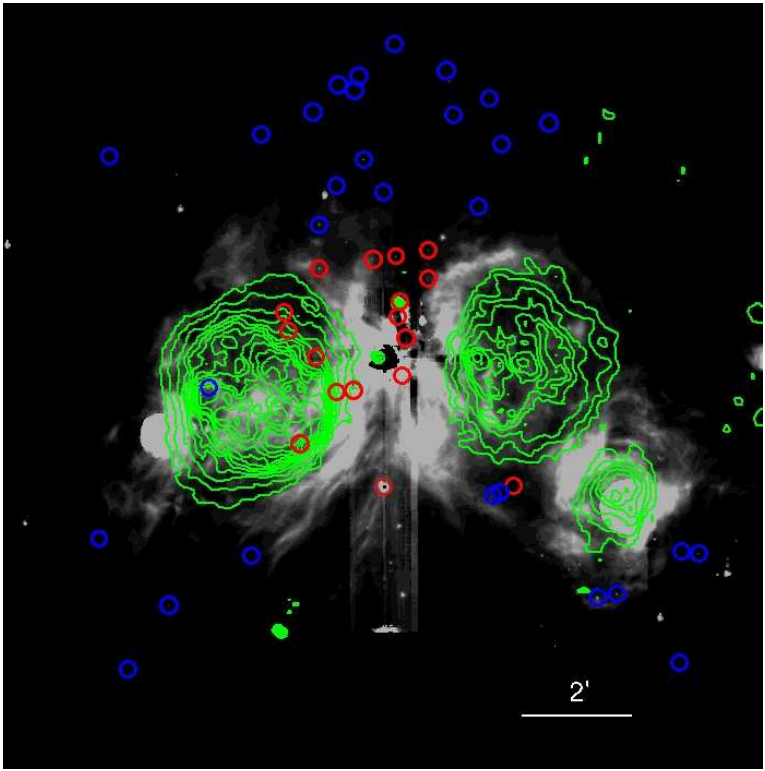}
% \vspace{1cm}
\caption{Spatial distribution of sources used for SED fitting, 
overlaid on the {\it Spitzer} IRAC
channel 4 (8.0 $\mu$m) image. 
The green contours represent the 610 MHz emissions taken from  Zinchenko et al. (in preparation). 
The blue circles represent sources with optical 
counterparts, whereas the red circles are the sources detected in NIR 
and longer wavelengths (see online electronic version for the color image).
}
\label{kdvel}
\end{figure*}
\end{document}